\documentclass[article, twocolumn,
   aps, pra,
  amsmath,amssymb,
  longbibliography,
  ]{revtex4-1}
\usepackage{graphicx,color}
\usepackage{amsmath}
\usepackage{natbib}
\usepackage{epsfig}
\begin{document}

\title{Elementary vibrational model for thermal conductivity of Lennard-Jones fluids: Applicability domain and accuracy level}

\author{S. A. Khrapak}\email{Sergey.Khrapak@gmx.de}
\affiliation{Joint Institute for High Temperatures, Russian Academy of Sciences, 125412 Moscow, Russia}
\author{A. G. Khrapak}
\affiliation{Joint Institute for High Temperatures, Russian Academy of Sciences, 125412 Moscow, Russia}

\begin{abstract}
Exact mechanisms of thermal conductivity in liquids are not well understood, despite rich research history. A vibrational model of energy transfer in dense simple liquids with soft pairwise interactions seems adequate to partially fill this gap. The purpose of the present paper is to define its applicability domain and to demonstrate how well it works within the identified applicability domain in the important case of the Lennard-Jones model system. The existing results from molecular dynamics simulations are used for this purpose. Additionally, we show that a freezing density scaling approach represents a very powerful tool to estimate the thermal conductivity coefficient across essentially the entire gas-liquid region of the phase diagram, including metastable regions. A simple practical expression serving this purpose is proposed.  
\end{abstract}

\date{\today}

\maketitle

\section{Introduction}

The thermal conductivity coefficient is an important characteristic of a material. It can vary greatly depending on specific substance as well on its phase state. While heat transport in gases, plasmas and solids is relatively well understood, mechanisms of heat transfer in liquids have remained elusive.   

A vibrational model of thermal conductivity in simple liquids with soft pairwise interactions has been recently proposed and discussed~\cite{KhrapakPRE01_2021}.
In this model it is assumed that atoms in liquids exhibit solid-like oscillations about temporary equilibrium positions corresponding to a local minimum on the system's potential energy surface~\cite{FrenkelBook,Stillinger1982,ZwanzigJCP1983}. The equilibrium positions are not fixed like in solids, but are allowed to diffuse (and this is why liquids can flow). However, this diffusion occurs on long enough time scales, which are irrelevant for the process of energy transfer. Furthermore, liquid is approximated by a quasi-layered structure with layers perpendicular to the temperature gradient and separated by the distance $\Delta = \rho^{-1/3}$, where $\rho$ is the atomic number density. The average interatomic separation in each quasi-layer is also $\Delta$. Then elementary consideration based on the assumption that the energy difference between the atoms in neighbouring layers is transferred at an average frequency of their solid-like vibrations leads immediately to a simple expression~\cite{KhrapakPRE01_2021}      
\begin{equation}\label{tc1}
\lambda\simeq c_{\rm v}\frac{\langle \omega \rangle}{2\pi \Delta}.
\end{equation} 
Here $\lambda$ is the thermal conductivity coefficient (which does not include the Boltzmann constant and hence is measured in cm$^{-1}$s$^{-1}$), $c_{\rm v}$ is the specific heat at constant volume, and $\langle \omega \rangle$ is the average vibrational frequency that approximates the energy transfer rate. Actually, specific heat at constant
volume $c_{\rm v}$ appears in Eq.~(\ref{tc1}) under the assumption of a small
thermal expansion coefficient. Dense liquids with soft interactions near the liquid-solid phase transition are essentially incompressible and this represents a good approximation.

We note in passing that the vibrational paradigm sketched above is useful not only in relation to the problem of thermal conductivity of dense fluids. It also allows to relate the self-diffusion and the shear viscosity coefficients in the form of Stokes-Einstein relation without the hydrodynamic diameter~\cite{ZwanzigJCP1983,CostigliolaJCP2019,KhrapakPRE10_2021,
KhrapakMolecules12_2021}. The present paper is, however, mostly focused on the thermal conductivity mechanism.   

Equation (\ref{tc1}) emphasises the relation between the thermal conductivity and collective modes properties of dense liquids. To evaluate the average vibrational frequency one would require to know the liquid vibrational density of states (VDOS). This is a formidable task, because collective properties can greatly differ from one liquid system to another and depend considerably on the state point in the phase diagram. Some progress has recently been reported~\cite{ZacconePNAS2021,StamperJPCL2022}. Zaccone and Baggioli have developed an analytical model for VDOS, based on overdamped Langevin liquid dynamics~\cite{ZacconePNAS2021}. Distinct from the Debye approximation, $g(\omega) \propto \omega^2$, for solids, the universal law for liquids reveals a linear relationship, $g(\omega)\propto \omega$, in the low-energy region. Stamper {\it et al.} have confirmed this universal law with experimental VDOS measured by inelastic neutron scattering on real liquid systems~\cite{StamperJPCL2022}. Nevertheless, the applicability regime and accuracy level of this model still require clarification. In the meantime, it is natural to employ some approximations and simplifications which can differ depending on the type of liquid under investigation (some examples are provided below). 

Interestingly enough, Eq.~(\ref{tc1}) can be reduced to the known previous results under special simplifying assumptions about the vibrational properties~\cite{KhrapakPRE01_2021}. In the simplest approximation all atoms are oscillating with the same Einstein frequency $\Omega_{\rm E}$ (this approximation is known as the Einstein model in the solid state physics). Averaging is then trivial, $\langle \omega \rangle=\Omega_{\rm E}$. We get
\begin{equation}\label{Horrocks}
\lambda \simeq c_{\rm v} \frac{\Omega_{\rm E}}{2\pi \Delta}.
\end{equation}   
This is very similar to to the results obtained by Rao~\cite{Rao1941} and later by Horrocks and McLaughlin~\cite{Horrocks1960,HorrocksTFS1963}. This form is particularly suitable for extremely soft interactions such as Coulomb and screened Coulomb potentials, relevant in the plasma-related context~\cite{KhrapakPoP01_2021,KhrapakPoP08_2021,KhrapakPPR2023}. In dense liquids near freezing conditions we can set $c_{\rm v}\sim 3$, according to Dulong-Petit law. The Einstein frequency at the liquid-solid phase transition can be estimated using the Lindeman's melting criterion in its simplest formulation.
Then, quasi-universality of the thermal conductivity coefficient at freezing conditions emerges in the form~\cite{Rao1941,Andrade1952}
\begin{equation}
\lambda \simeq {\rm const} \sqrt{\frac{T}{m\Delta^4}},
\end{equation}     
where $T$ is the temperature in energy units ($=k_{\rm B}T$) and $m$ is the atomic mass. The value of the constant is about $\simeq 10$ for usual simple model systems such as Lennard-Jones, Coulomb, screened Coulomb, as well as monatomic liquids (e.g. liquefied noble gases)~\cite{KhrapakPoF2022,KhrapakJCP2022_1}. It is somewhat larger for the hard-sphere fluid at freezing and reaches values $\simeq 15 - 20$ for liquids with more complex molecular structure~\cite{KhrapakJMolLiq2023}.   

Alternatively, we can relate the average frequency to the sound velocity. Namely, consider an acoustic dispersion of the form $\omega=kc_{\rm s}$, where $k$ is the wave-number and $c_{\rm s}$ is the sound velocity. The characteristic wave-number for energy transfer between nearest neighbours is $k\sim 2\pi/\Delta$. This yields $\langle \omega\rangle \sim 2\pi c_{\rm s}/\Delta$ and hence
\begin{equation}\label{Bridgman}
\lambda\sim c_{\rm v}\frac{c_{\rm s}}{\Delta^2}.
\end{equation}        
This resembles Bridgman's expression for the thermal conductivity coefficient~\cite{Bridgman1923}. Actually, Bridgman postulated that the energy is transferred at the speed of sound, thus leading to a linear correlation between the thermal conductivity coefficient and the sound velocity. He also used a constant coefficient of two, that is $\lambda\simeq 2c_{\rm s}/\Delta^2$. This numerical coefficient remained somewhat controversial, values between 2 and 3 were used in the literature~\cite{ZhaoJAP2021,XiCPL2020,KhrapakPRE01_2021,BirdBook}.
In a recent extensive study of correlations between the thermal conductivity coefficient and the sound velocity of various liquids~\cite{KhrapakJMolLiq2023,KhrapakPoF2023}, it has been demonstrated that linear correlations are well reproduced for model liquids as well as real monatomic and diatomic liquids. However, they are less convincing in polyatomic molecular liquids. The actual coefficient of proportionality in Bridgman's formula is not fixed. It is about unity for monatomic liquids and generally increases with molecular complexity.

Just like in solids, dense liquids support one longitudinal and two transverse collective modes~\cite{HansenBook,BalucaniBook,OhtaPRL2000,HosokawaJPCM2015,BrykJCP2017}.   A Debye-like approximation for the vibrational spectra of dense liquid is thus not completely irrelevant and has demonstrated reasonable success in deriving the Stokes-Einstein coefficient without the hydrodynamic diameter~\cite{ZwanzigJCP1983} and estimating thermodynamic properties of dense liquids ~\cite{BolmatovSciRep2012,KhrapakJCP2021}. In this approach the dispersion relations of one longitudinal and two transverse collective modes are approximated by their corresponding acoustic asymptotes $\omega_{\rm l}(k)=kc_{\rm l}$ and $\omega_{\rm t}(k)=kc_{\rm t}$, terminating at respective cut-off wavelengths $k_{\rm cut}$. Application to the problem of thermal conductivity results in the expression~\cite{KhrapakPRE01_2021}
\begin{equation}\label{Cahill1}
\lambda \simeq \frac{1}{4}\left(\frac{3}{4\pi}\right)^{1/3} c_{\rm v} \frac{c_l+2c_t}{\Delta^2}. 
\end{equation}
Using $c_{\rm v}\simeq 3$ near freezing we get a formula similar to that of minimal thermal conductivity model proposed by Cahill and Pohl~\cite{Cahill1989,Cahill1992} for amorphous solids. 

Equation (\ref{Cahill1}) can be applicable to liquids with conventional acoustic dispersion relation of the longitudinal collective mode, but should not be used when dispersion is non-acoustic (e.g. in the case of Coulomb one-component plasma and weakly screened Yukawa systems). In such cases actual dispersion relations can be used to perform averaging~\cite{KhrapakPoP01_2021,KhrapakMolecules12_2021}). Another complication arises due to the presence of the so called $k$-gap (zero-frequency domain at sufficiently long wavelengths) in the dispersion relation of the liquid transverse mode, which has received considerable attention in recent years~\cite{GoreePRE2012,YangPRL2017,TrachenkoRPP2015,KhrapakJCP2019,
KryuchkovSciRep2019,KryuchkovJCP2021}. The effect of $k$-gap in the transverse mode has not yet been discussed in detail in the context of the vibrational paradigm of transport properties in liquids and will not be addressed here.     
   
Formula (\ref{Cahill1}) provides a good compromise between simplicity and accuracy. Notably, it does not contain free parameters and adjustable coefficients. 

One of us demonstrated previously that Eq.~(\ref{Cahill1}) applies reasonably well to dense Lennard-Jones (LJ) fluids~\cite{KhrapakPRE01_2021}. However, only a single exemplary slightly supercritical isotherm was considered in that work. The purpose of this paper is to provide a detailed and extensive verification of its applicability domain and accuracy level across the LJ fluid phase diagram. We demonstrate that for the LJ fluid, Eq.~(\ref{Cahill1}) allows to express the thermal conductivity coefficient in terms of thermodynamic properties such as specific heat, excess internal energy, and excess pressure. An appropriate equation of state is then used to evaluate the thermal conductivity coefficient and to quantify the accuracy of the approach. This is a remarkable example of a direct relation between transport and thermodynamics. The applicability domain is defined with reference to a gas-to-liquid dynamical crossover in supercritical fluids (Frenkel line in the phase diagram), which represents an important current research topic. Finally, we reiterate our recent freezing density scaling approach to transport properties~\cite{KhrapakPRE04_2021,KhrapakJPCL2022,KhrapakJCP2022_1}, and put forward an {\it ad hoc} expression for the thermal conductivity coefficient. Overall, the results reported represent an important step toward better understanding main mechanisms and peculiarities of transport phenomena in the liquid state.      

\section{Methods}

The LJ model is one of the most popular and best studied systems in condensed matter research. It combines relative simplicity with adequate representation (at least at the qualitative level) of interatomic interactions in real substances, exhibiting steep short-range repulsion and softer long-range attraction. The LJ potential is 
\begin{equation}
\phi(r)=4\epsilon\left[\left(\frac{\sigma}{r}\right)^{12}-\left(\frac{\sigma}{r}\right)^{6}\right], 
\end{equation}
where  $\epsilon$ and $\sigma$ are the energy and length scales (or LJ units). The reduced density and temperature expressed in LJ units are $\rho^*=\rho\sigma^3$, $T^*=T/\epsilon$.  

Transport properties of the LJ system have been extensively studied in the literature. Recent overview of available simulation data can be found in  Refs.~\cite{BellJPCB2019,HarrisJCP2020,AllersJCP2020}. Particularly extensive and useful datasets have been published by Meier {\it et al.}~\cite{Meier2002,MeierJCP_1,MeierJCP_2} and by Baidakov {\it et al.}~\cite{BaidakovFPE2011,BaidakovJCP2012,BaidakovJCP2014}. 
Transport data have been tabulated along different isotherms in a wide regions of the LJ system phase diagram. Good agreement between the two datasets for overlapping regimes has been reported~\cite{HarrisJCP2020}. 
Previously, we used the numerical results of Meier~\cite{Meier2002} to demonstrate the adequacy of Eq.~(\ref{Cahill1}). Using the data for the thermal conductivity coefficient, specific heat, reduced energy and pressure along a single close-critical isotherm $T^* = 1.35$ tabulated in Ref.~\cite{MeierJCP_1} we documented a very good accuracy of Eq.~(\ref{Cahill1}) in the dense liquid regime. Here we first make use of the thermal conductivity data from Ref.~\cite{BaidakovJCP2014}, which cover a rather extended area on the LJ system phase diagram.

\begin{figure}
\includegraphics[width=8.5cm]{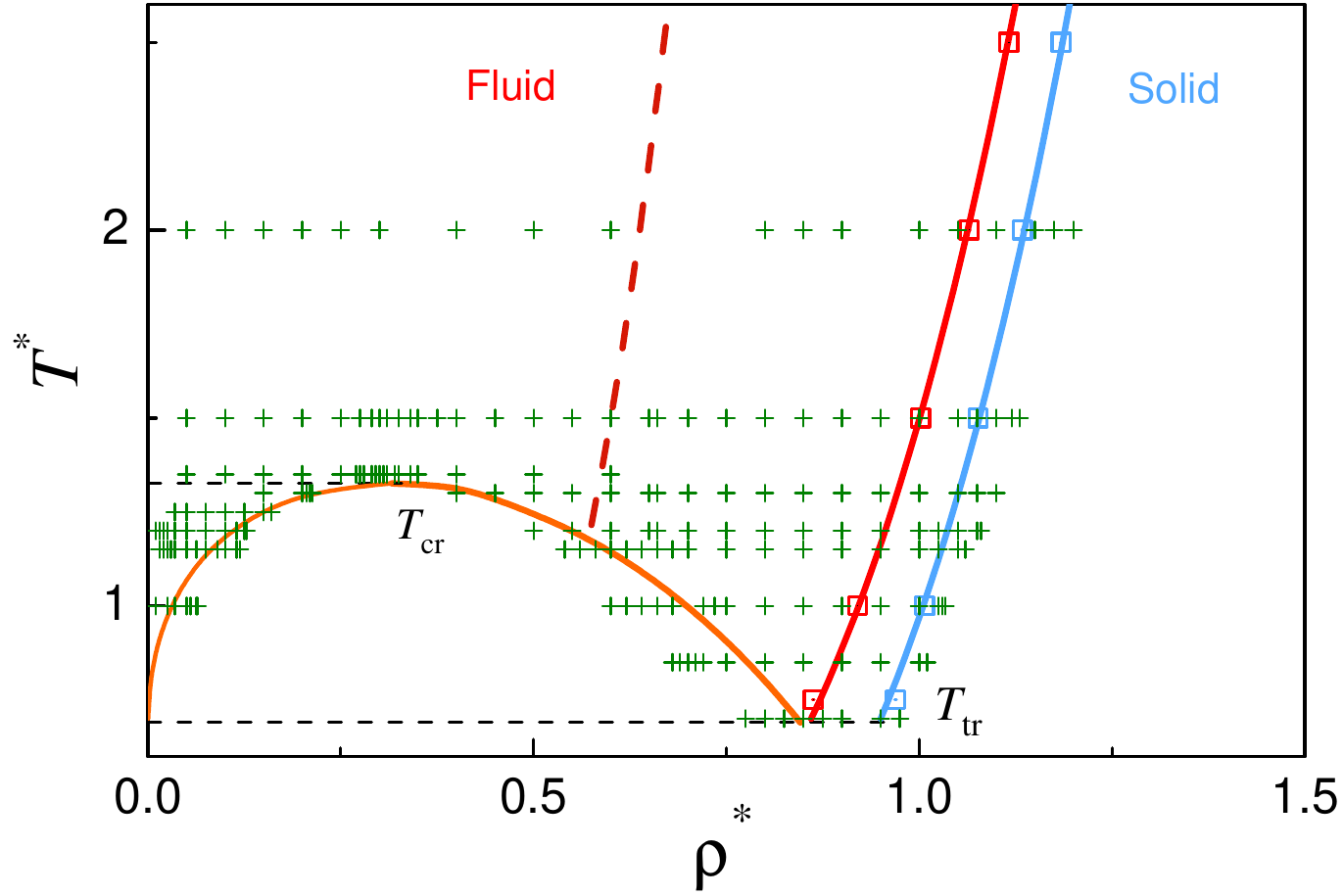}
\caption{(Color online) Phase diagram of the LJ system on the density-temperature plane ($\rho^*$, $T^*$). The squares  correspond to the fluid-solid coexistence boundaries as tabulated in Refs.~\cite{SousaJCP2012}, the fluid-solid coexistence curves are simple fits by Eq.~(\ref{fit_freezing}). The liquid-vapour coexistence boundary is plotted using the formulas provided in Ref.~\cite{HeyesJCP2019}. 
The reduced triple point and critical temperatures are $T^*_{\rm tr}\simeq 0.69$~\cite{SousaJCP2012} and $T^*_{\rm c}\simeq 1.33$~\cite{HeyesJCP2019}, respectively. The crosses correspond to the state points investigated in Ref.~\cite{BaidakovJCP2014}. Only data points with the temperature above the triple point are considered. The dashed line corresponding to a fixed reduced density $\rho/\rho_{\rm fr}\simeq 0.6$ marks the onset of the vibrational model applicability (for further details see the text). 
}
\label{FigLJ_PD}
\end{figure}

The phase diagram of LJ system is shown in Fig.~\ref{FigLJ_PD}. Here the solid-liquid coexistence data are taken from Ref.~\cite{SousaJCP2012}. For convenience, they are fitted by simple expressions
\begin{equation}\label{fit_freezing}
\begin{aligned}
T^*_{\rm fr}&\simeq 2.111(\rho^*)^4-0.615(\rho^*)^2, \\
T^*_{\rm m}&\simeq 1.988(\rho^*)^4-1.019(\rho^*)^2,
\end{aligned}
\end{equation}
where subscripts ``${\rm fr}$'' and ``${\rm m}$'' denote freezing and melting. respectively. The functional form of these expressions arises in various simple approaches to fluid-solid coexistence in the LJ model~\cite{RosenfeldCPL1976,RosenfeldMolPhys1976,KhrapakPRB2010,
KhrapakJCP2011_2,PedersenNatCom2016,CostigliolaPCCP2016}.
The liquid-vapour coexistence boundary is plotted using the formulas provided in Ref.~\cite{HeyesJCP2019}. The crosses are the state points for which the thermal conductivity coefficients was numerically evaluated by means of equilibrium molecular dynamics with the use of the Green-Kubo formalism in Ref.~\cite{BaidakovJCP2014}. Some of investigated state points correspond to metastable regions: superheated and supercooled liquids, supersaturated vapour. Here we only consider state points with the temperature above the triple point temperature (lower temperatures are not relevant because they are beyond the applicability limits of the vibrational model). An additional dashed line marks the onset of validity of the vibrational model. This was identified in Ref.~\cite{KhrapakPRE10_2021} from the analysis of the Stokes-Einstein (SE) product $D\eta(\Delta/T)$, where $D$ is the self-diffusion coefficient and $\eta$ is the shear viscosity coefficient. At low densities the SE product scales as $(\rho^*)^{-4/3}$ as it should in the gaseous regime. However, it approaches a constant asymptotic value of $\simeq 0.15$ at sufficiently high densities, not too far from that at freezing. The two asymptotes intersect at $\rho/\rho_{\rm fr}\simeq 0.35$ and this can be considered as an indication of the gas-to-fluid dynamical crossover~\cite{KhrapakJCP2022}. At about twice this density,    $\rho/\rho_{\rm fr}\gtrsim 0.6$, the SE product becomes practically constant and this can be identified as a lower boundary of the validity of the SE relation and hence of the vibrational picture of atomic dynamics~\cite{KhrapakPRE10_2021}. Note, that the analysis of various model systems (such as Lennard-Jones, Coulomb, Yukawa) indicated that the onset of validity of vibrational dynamics corresponds to nearly the same value of excess entropy, $s_{\rm ex}\simeq -2$~\cite{KhrapakPRE10_2021}, and this can be considered as a more general condition of the validity of the vibrational model.        

The numerical results for the thermal conductivity coefficient are traditionally expressed in LJ units as
\begin{equation}\label{LJunits}
\lambda^*=\frac{\lambda\sigma^2}{k_{\rm B}}\left(\frac{m}{\epsilon}\right)^{1/2}.
\end{equation} 
Since we measure temperature in energy units, the Boltzmann constant $k_{\rm B}$ should disappear. Also, we find it more convenient to work with macroscopically reduced units (sometimes referred to as Rosenfeld's normalization~\cite{RosenfeldJPCM1999}). This is particularly advantageous when comparing transport properties of different systems~\cite{RosenfeldPRA1977,RosenfeldJPCM1999,KhrapakPR2023}. The macroscopically reduced thermal conductivity coefficient reads
\begin{equation}\label{Rnorm}
\lambda_{\rm R}=\lambda\rho^{-2/3}\left(\frac{m}{T}\right)^{1/2}.
\end{equation}      
From Eqs.~(\ref{LJunits}) and (\ref{Rnorm}) a trivial relation emerges
\begin{equation}\label{Relation}
\lambda_{\rm R}=\frac{\lambda^*}{(T^*)^{1/2}(\rho^*)^{2/3}}. 
\end{equation}
This relation has been used to evaluate $\lambda_{\rm R}$ from the available numerical results.  

To compare with the vibrational mechanism of heat transfer, the longitudinal and transverse instantaneous sound velocities have to be evaluated. For the LJ system these can be expressed using the excess energy and pressure of the system as demonstrated by Zwanzig and Mountain~\cite{ZwanzigJCP1965} (see also Ref.~\cite{KhrapakMolecules2020} for further details): 
\begin{equation}\label{cl}
\frac{c_l^2}{v_{\rm T}^2}=3-\frac{72}{5}u_{\rm ex}+11p_{\rm ex},
\end{equation} 
\begin{equation}\label{ct}
\frac{c_t^2}{v_{\rm T}^2}=1-\frac{24}{5}u_{\rm ex}+3p_{\rm ex}.
\end{equation}
Here $u_{\rm ex}=U/NT-3/2$ is the excess internal energy, $p_{\rm ex}=P/\rho T-1$ is the excess pressure, $U$, $P$, and $N$ being the internal energy,  pressure, and the number of atoms, respectively. The sound velocities are expressed in units of the thermal velocity $v_{\rm T}=\sqrt{T/m}$. It is a simple exercise to verify that Cauchy relation is satisfied in this approach:
\begin{equation}
\left(\frac{c_l}{v_{\rm T}}\right)^2-3\left(\frac{c_t}{v_{\rm T}}\right)^2=2p_{\rm ex}.
\end{equation} 
   
In order to evaluate the sound velocities, an appropriate equation of state (EoS) is required. 
A large number of different EOS of have been proposed in the literature to describe thermodynamic properties of LJ fluids (see e.g. Refs. \cite{Stephan2019,Stephan2020} for review). Here we use the equation of state developed by Thol {\it et al}.~\cite{Thol2016}. This is an empirical equation of state, formulated in terms of the Helmholtz free energy and based on a large molecular simulation data set and thermal virial coefficients. Its applicability range $0.7< T^*<9$, $\rho^*<1.08$ is sufficient for the present purpose. This EoS demonstrates reasonable level of accuracy when compared to others~\cite{Stephan2020}, is relatively simple and convenient in implementation. We use Thol {\it et al}. EoS to calculate the specific heat $c_{\rm v}$, excess pressure $p_{\rm ex}$ and energy $u_{\rm ex}$. From the latter thermodynamic quantities the sound velocities $c_l$ and $c_t$ are calculated. This is all input needed to calculate the thermal conductivity coefficient from Eq.~(\ref{Cahill1}). Thus, application of the vibrational model to LJ fluids represents a remarkable example when a transport property (thermal conductivity in the considered case) is determined explicitly by thermodynamic quantities and knowledge of EoS is sufficient to calculate it.    
In Section~\ref{Results} we will see how well theory compares with results from numerical simulations. 

\begin{figure*}
\includegraphics[width=17cm]{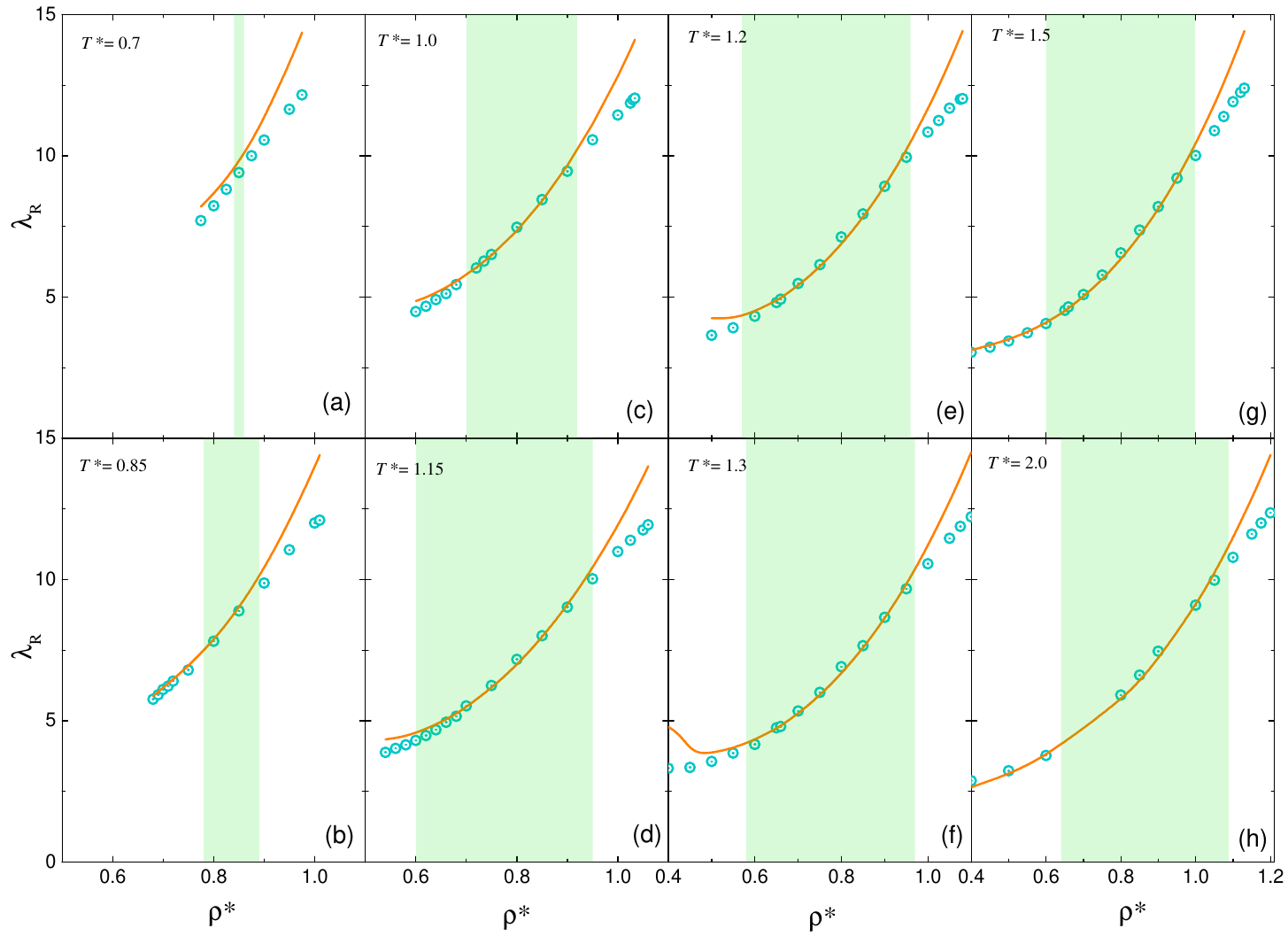}
\caption{(Color online) Reduced thermal conductivity coefficient $\lambda_{\rm R}$ of the Lennard-Jones fluid as a function of the reduced density $\rho^*$. Results for eight isotherms are shown: $T^*=0.7$ (a), 0.85 (b), 1.0(c), 1.15 (d), 1.2 (e), 1.3 (f), 1.5 (g), and 2.0 (h). The circles denote the numerical results from Ref.~\cite{BaidakovJCP2014}, the solid curves are calculated by means of vibrational model using Eq.~(\ref{Cahill1}), as discussed in the text. The shaded regions in each figure mark the conditions of applicability of the theory. }
\label{Compar}
\end{figure*} 

\section{Results}\label{Results}

Figure~\ref{Compar} demonstrates the comparison between the vibrational model, as calculated using Eq.~(\ref{Cahill1}) complemented by the thermodynamic quantities from Thol {\it et al}. EoS~\cite{Thol2016}, with the MD results from Baidakov {\it et al}.~\cite{BaidakovJCP2014}. The circles correspond to the results from MD simulation. We have selected only the isotherms with $T^*>T^*_{\rm tp}$, where vibrational model makes sense and Thol~{\it et al}. EoS is applicable. We have also considered supercritical densities for which the vibrational model might be adequate. For this reason we, have omitted the two isotherms $T^*=1.25$ and $T^*=1.35$ for which only irrelevant low-density data are available. The remaining data for eight isotherms are depicted in Fig.~\ref{Compar} (a)-(h). The solid curves in each figure correspond to our theoretical calculation.

The shaded regions in each figure correspond to the regime of applicability of the present model. From the side of high densities the applicability is limited by the fluid boundary of the fluid-solid coexistence. From the side of low densities the onset of the applicability is either the condition $\rho/\rho_{\rm fr}\simeq 0.6$ where vibrational picture is adequate or the liquid boundary of the gas-liquid coexistence. Although metastable state points are present in Fig.~\ref{Compar} they should not be used to judge the applicability of the vibrational model. Neither the model itself was designed to deal with coexisting phases, nor the EoS from Ref.~\cite{Thol2016} can be expected to be reliable there.

\begin{figure}
\includegraphics[width=8cm]{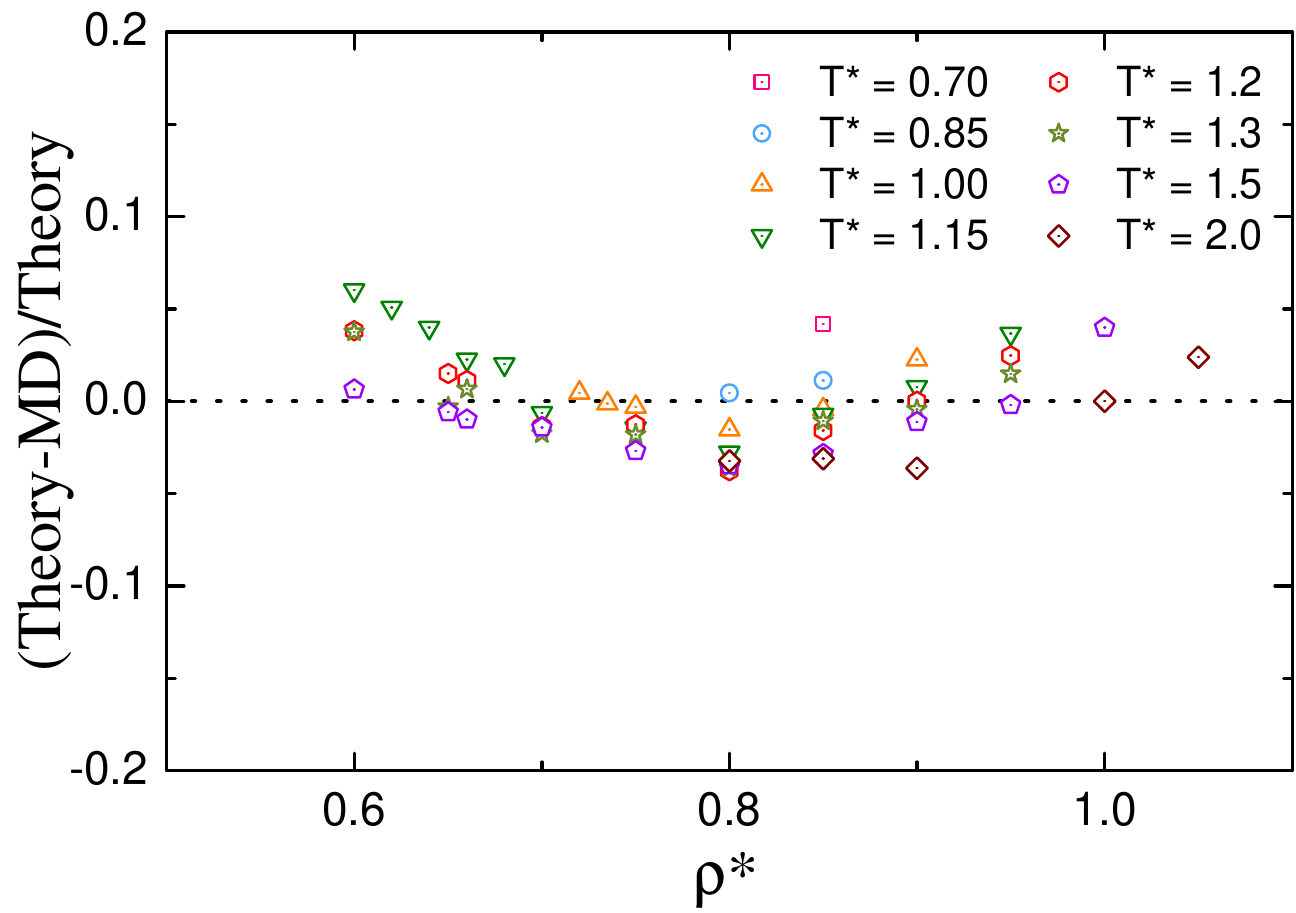}
\caption{(Color online) Relative disagreement between theory [Eq.~(\ref{Cahill1})] and MD simulations of Ref.~\cite{BaidakovJCP2014}. The quantity $(\lambda_{\rm Theory}-\lambda_{\rm MD})/\lambda_{\rm Theory}$ is plotted as a function of the reduced density $\rho^*$ for eight isotherms. Only the state points within the applicability regime of the theoretical model are considered.      }
\label{Accuracy}
\end{figure} 

Careful examination of eight panels in Fig.~\ref{Compar} demonstrates a very good agreement between the theoretical calculation and MD simulation in the regime where theory is applicable. This is further quantified in Fig.~\ref{Accuracy}. Relative deviations are typically limited by few percent, increasing to $\simeq 5\%$ towards the boundaries of the applicability domain. Some simulation data for the thermal conductivity reported previously have considerably larger 
standard deviations compared to the observed deviations (see e.g. Table II in Ref.~\cite{NasrabadJCP2006}). For this reason the agreement can be considered as excellent, especially taking into account that the theoretical model does not contain any free adjustable parameters. 

\begin{figure}
\includegraphics[width=8cm]{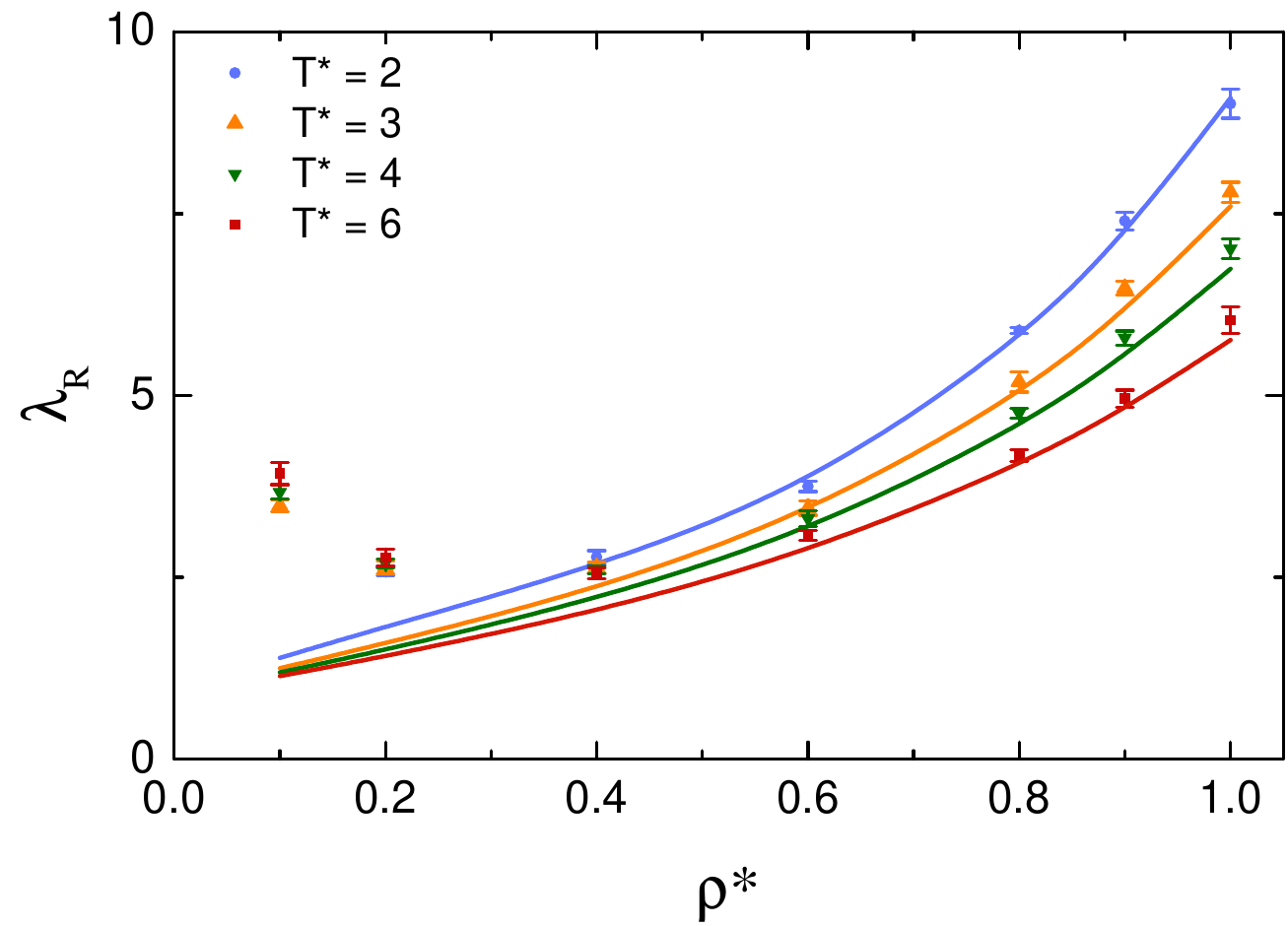}
\caption{(Color online) Reduced thermal conductivity coefficient $\lambda_{\rm R}$ of the Lennard-Jones fluid as a function of the reduced density $\rho^*$. The symbols denote the numerical results from the non-equilibrium simulations in Ref.~\cite{GallieroPRE2009}. The solid curves are calculated using Eq.~(\ref{Cahill1}).}
\label{Galliero}
\end{figure}

To reinforce this conclusion and demonstrate that the documented agreement is not the result of some fortunate coincidence, we have performed an additional comparison using the set of simulation data reported by Galliero and Boned~\cite{GallieroPRE2009}. The purpose is two-fold. First, in contrast to Ref.~\cite{BaidakovJCP2014}, Galliero and Boned employed {\it non-equilibrium} molecular dynamics simulations to obtain thermal conductivity of LJ fluids. In this way, we can potentially quantify discrepancies between different methods, if such exist. Second, the data set from Ref.~\cite{GallieroPRE2009} covers considerably higher range of temperatures and hence we can verify whether the vibrational model remains meaningful as the temperature further increases. Figure~\ref{Galliero} presents the comparison between the theory and non-equilibrium simulations. We observe that for $\rho^*\gtrsim 0.6$ the deviation between theory and numerical experiment is mostly within numerical data uncertainty. For a common temperature $T^*=2$ the theoretical model compares equally well with equilibrium and non-equilibrium simulations, demonstrating no major discrepancies between the two latter. For higher temperatures there is no any sign that the theory may behave inappropriately.  

Thus, where applicable the vibrational model of heat transport agrees very well with the results from different numerical simulations. Yet, there is another practical approach -- the freezing density scaling (FDS) -- which allows to predict the transport properties, including the thermal conductivity coefficient, in an even wider parameter regime. We would like to use this opportunity and reiterate the application of FDS scaling to the thermal conductivity coefficient of the LJ fluid. Section~\ref{SecFDS} serves this purpose.

\section{Freezing density scaling}\label{SecFDS}

It has been recently demonstrated that macroscopically reduced self-diffusion, shear viscosity, and thermal conductivity coefficients of LJ fluids along isotherms exhibit quasi-universal scaling on the density divided by its value at the freezing point, $\rho_{\rm fr}$~\cite{KhrapakPRE04_2021}. 
Originally considered as a useful empirical observation~\cite{KhrapakPRE04_2021}, it has been later discussed in the context of   
quasi-universal excess entropy scaling and isomorph theory~\cite{KhrapakJPCL2022,KhrapakJCP2022_1,HeyesJCP2023}. 
FDS approach implies that the transport coefficients are functions of a single variable $\rho/\rho_{\rm fr}$, and thus it represents a very convenient corresponding states principle to estimate transport properties in LJ fluids.
Importantly, it has been shown that FDS holds even at quite low densities where neither the original form of the excess entropy scaling nor the isomorph theory are expected to work.  Additionally, the functional form of the FDS scaling is similar (although not identical) to that in the hard-sphere fluid. Thus, it can be expected that FDS is not a special property of LJ fluids, but applies (possibly with some modifications) to a wider class of fluids.   

\begin{figure}
\includegraphics[width=8cm]{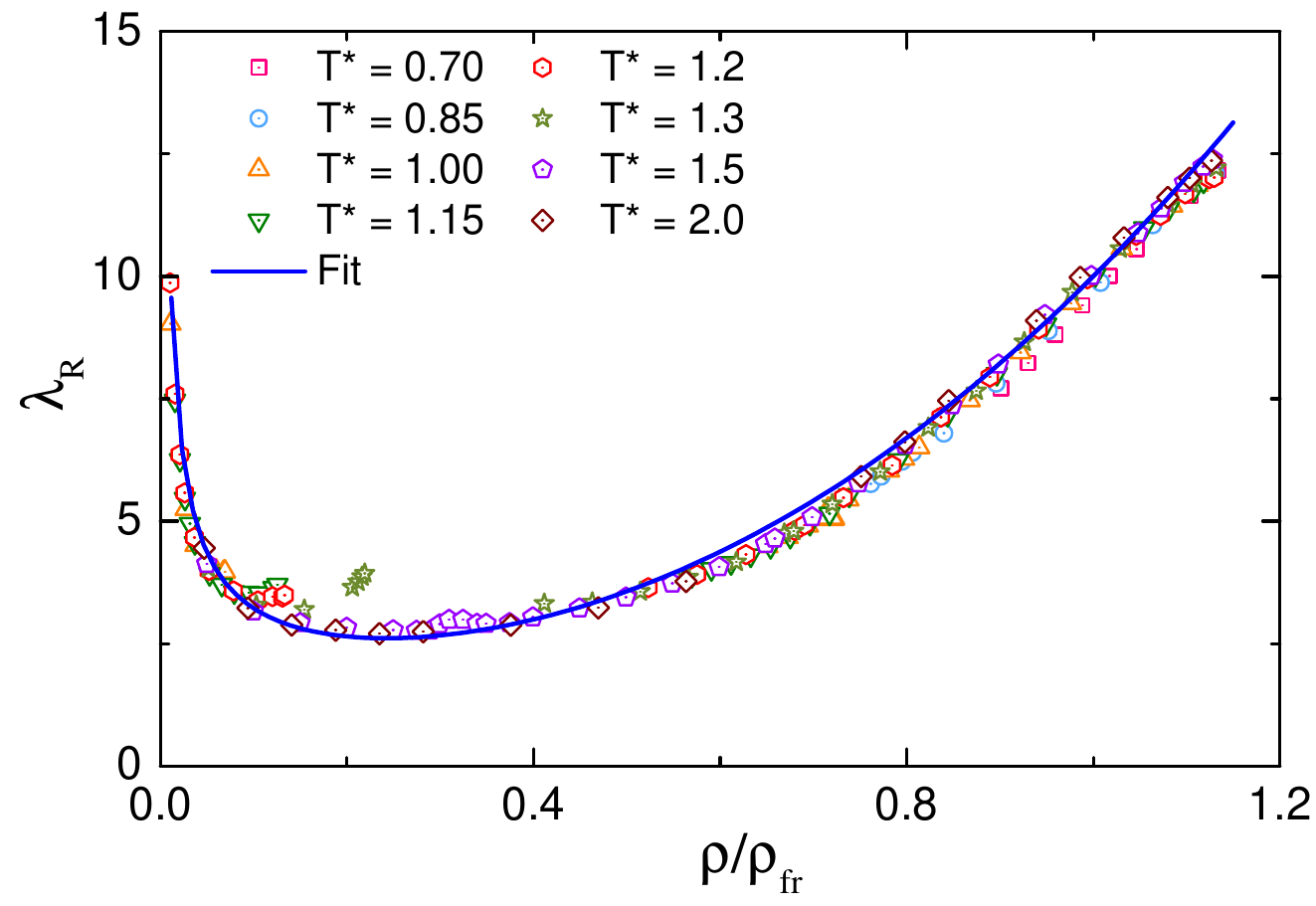}
\caption{(Color online) Reduced thermal conductivity coefficient $\lambda_{\rm R}$ vs reduced density $\rho/\rho_{\rm fr}$ demonstrating the success of the freezing density scaling of transport coefficients. Symbols correspond to the numerical data from Baidakov {\it et al}~\cite{BaidakovJCP2014}. along different isotherms (see the label). The solid curve corresponds to a quasi-universal fit of Eq.~(\ref{fiteq}).}
\label{FDS}
\end{figure}   
  
Detailed analysis of the numerical dataset from Ref.~\cite{BaidakovJCP2014} has been further extended to reinforce the FDS concept. The results are shown in Fig.~\ref{FDS}, demonstrating that the data tend to collapse on a universal master curve. The only region where the FDS fails is the vicinity of the critical point, where critical enhancement of the thermal conductivity coefficient becomes important. 
For the dataset produced in Ref.~\cite{BaidakovJCP2014} this concerns only a few data points, mostly on the near-critical isotherm $T^*=1.3$ (see Fig.~\ref{FDS}). Importantly, FDS approach seemingly works well for metastable state points, which are also shown in Fig.~\ref{FDS}. No clear distinction between thermodynamically stable and metastable states can be seen. Also, it is remarkable that the FDS approach applies to rather low densities. This is rather unexpected, because at such low densities it is natural to assume that any ''memory'' about the location of the fluid-solid phase transition and the freezing density is completely lost. However, it is necessary to remark that in the low-density gaseous regime established methods to calculate the transport cross sections and the transport properties of the LJ system do exist (see e.g. Refs.~\cite{HirschfelderBook,HirschfelderJCP1948,SmithJCP1964,
KhrapakPRE2014_scattering,KhrapakEPJD2014,KimJCompPhys2014,
Kristiansen2020} and references therein for some related works). 

The appealing quasi-universality of the data shown in Fig.~\ref{FDS} calls for an appropriate mathematical description. 
Among various fitting functions that we attempted, a particularly simple but appropriate form is provided by 
\begin{equation}\label{fiteq}
\lambda_{\rm R}=\frac{\alpha}{{\mathcal R}^{2/3}}+\beta +\gamma{\mathcal R}^{\delta}.
\end{equation}
Here ${\mathcal R}$ is the density ratio, ${\mathcal R}=\rho/\rho_{\rm fr}$. The first term is chosen as to provide a correct asymptote of the thermal conductivity coefficient in the low-density limit, $\lambda_{\rm R}\propto (\rho^*)^{-2/3}$~\cite{KhrapakPRE04_2021}.
Furthermore, $\alpha$, $\beta$, $\gamma$, and $\delta$ are the fitting coefficients. Based on the dataset for the supercritical isotherm $T^*=2$, the coefficients $\alpha\simeq 0.43$, $\beta\simeq 1.18$, $\gamma\simeq 8.39$, and $\delta\simeq 2.30$ have been obtained. The solid curve in Fig.~\ref{FDS} represents this fit. It is representative for other isotherms, except those very near the critical temperature, because the critical enhancement is not accounted for. Importantly, it appears also representative for all metastable state points that were investigated in Ref.~\cite{BaidakovJCP2014}.   

\section{Conclusion}

Thermal conductivity is an important characteristic of a material. While this property is relatively well understood in gases, solids and plasmas, the same cannot be said about liquids. In this paper we have considered the Lennard-Jones fluid as an important simple model system to investigate the applicability limits and check the accuracy of theoretical approaches to the thermal conductivity coefficient. This program has been realised with the help of an extensive MD simulation datasets provided by Baidakov {\it et al}.~\cite{BaidakovJCP2014} and Galliero and Boned~\cite{GallieroPRE2009}.     

Two theoretical methods have been highlighted in this paper. The first one -- the vibrational paradigm of heat transfer -- leads to an expression [Eq.~(\ref{Cahill1})], which relates the thermal conductivity coefficient to the thermodynamic properties of the LJ fluid. Although very simplistic and to some extent naive, the vibrational model is able to describe the numerical data on the thermal conductivity coefficient with a remarkable accuracy. No adjustable parameters are involved and the theoretical calculation requires only the input from thermodynamics. The applicability region spans from the gas-liquid coexistence boundary to the liquid-solid coexistence boundary for subcritical temperatures and from a constant density ratio $\rho/\rho_{\rm fr}\simeq 0.6$ to the fluid-solid coexistence boundary for supercritical temperatures. 

The second method is the corresponding state principle, based on the freezing density scaling approach. This is more intuitive approach, although solid relations to the excess entropy scaling and the isomorph theory do exist. The FDS approach implies that the reduced thermal conductivity coefficients depends quasi-universally on the density reduced by its value at the freezing point (at a given temperature). The applicability domain of the FDS approach covers almost the entire gas-liquid-fluid region in the LJ system phase diagram (probably excluding very low densities and deep gas-liquid coexistence region). It is applicable even to metastable
regions of superheated and supercooled liquids and supersaturated vapour. 
This can make the FDS approach a very useful practical tool to estimate the thermal conductivity coefficient under various conditions and a tentative fitting formula is provided for this purpose. The only domain where FDS clearly fails is the close vicinity of the critical point, because it does not account for critical enhancement.                 

Overall, the reported results shed new light on the properties and mechanisms of heat transport in liquids and provide new accurate methods for the estimation of the thermal conductivity coefficient. This can be of interest for researchers in condensed matter, physics of fluids, materials science and beyond.    

The authors have no conflicts of interest to disclose.

Data sharing is not applicable to this article as no new data were created or analyzed in this study. 




\bibliography{SE_Ref}

\providecommand{\noopsort}[1]{}\providecommand{\singleletter}[1]{#1}%
\begin{thebibliography}{78}%
\makeatletter
\providecommand \@ifxundefined [1]{%
 \@ifx{#1\undefined}
}%
\providecommand \@ifnum [1]{%
 \ifnum #1\expandafter \@firstoftwo
 \else \expandafter \@secondoftwo
 \fi
}%
\providecommand \@ifx [1]{%
 \ifx #1\expandafter \@firstoftwo
 \else \expandafter \@secondoftwo
 \fi
}%
\providecommand \natexlab [1]{#1}%
\providecommand \enquote  [1]{``#1''}%
\providecommand \bibnamefont  [1]{#1}%
\providecommand \bibfnamefont [1]{#1}%
\providecommand \citenamefont [1]{#1}%
\providecommand \href@noop [0]{\@secondoftwo}%
\providecommand \href [0]{\begingroup \@sanitize@url \@href}%
\providecommand \@href[1]{\@@startlink{#1}\@@href}%
\providecommand \@@href[1]{\endgroup#1\@@endlink}%
\providecommand \@sanitize@url [0]{\catcode `\\12\catcode `\$12\catcode
  `\&12\catcode `\#12\catcode `\^12\catcode `\_12\catcode `\%12\relax}%
\providecommand \@@startlink[1]{}%
\providecommand \@@endlink[0]{}%
\providecommand \url  [0]{\begingroup\@sanitize@url \@url }%
\providecommand \@url [1]{\endgroup\@href {#1}{\urlprefix }}%
\providecommand \urlprefix  [0]{URL }%
\providecommand \Eprint [0]{\href }%
\providecommand \doibase [0]{http://dx.doi.org/}%
\providecommand \selectlanguage [0]{\@gobble}%
\providecommand \bibinfo  [0]{\@secondoftwo}%
\providecommand \bibfield  [0]{\@secondoftwo}%
\providecommand \translation [1]{[#1]}%
\providecommand \BibitemOpen [0]{}%
\providecommand \bibitemStop [0]{}%
\providecommand \bibitemNoStop [0]{.\EOS\space}%
\providecommand \EOS [0]{\spacefactor3000\relax}%
\providecommand \BibitemShut  [1]{\csname bibitem#1\endcsname}%
\let\auto@bib@innerbib\@empty
\bibitem [{\citenamefont {Khrapak}(2021{\natexlab{a}})}]{KhrapakPRE01_2021}%
  \BibitemOpen
  \bibfield  {author} {\bibinfo {author} {\bibfnamefont {S.~A.}\ \bibnamefont
  {Khrapak}},\ }\bibfield  {title} {\enquote {\bibinfo {title} {Vibrational
  model of thermal conduction for fluids with soft interactions},}\ }\href
  {\doibase 10.1103/physreve.103.013207} {\bibfield  {journal} {\bibinfo
  {journal} {Phys. Rev. E}\ }\textbf {\bibinfo {volume} {103}},\ \bibinfo
  {pages} {013207} (\bibinfo {year} {2021}{\natexlab{a}})}\BibitemShut
  {NoStop}%
\bibitem [{\citenamefont {Frenkel}(1955)}]{FrenkelBook}%
  \BibitemOpen
  \bibfield  {author} {\bibinfo {author} {\bibfnamefont {Y.}~\bibnamefont
  {Frenkel}},\ }\href {https://cds.cern.ch/record/106808} {\emph {\bibinfo
  {title} {{Kinetic theory of liquids}}}}\ (\bibinfo  {publisher} {Dover},\
  \bibinfo {address} {New York, NY},\ \bibinfo {year} {1955})\BibitemShut
  {NoStop}%
\bibitem [{\citenamefont {Stillinger}\ and\ \citenamefont
  {Weber}(1982)}]{Stillinger1982}%
  \BibitemOpen
  \bibfield  {author} {\bibinfo {author} {\bibfnamefont {F.~H.}\ \bibnamefont
  {Stillinger}}\ and\ \bibinfo {author} {\bibfnamefont {T.~A.}\ \bibnamefont
  {Weber}},\ }\bibfield  {title} {\enquote {\bibinfo {title} {Hidden structure
  in liquids},}\ }\href {\doibase 10.1103/physreva.25.978} {\bibfield
  {journal} {\bibinfo  {journal} {Phys. Rev. A}\ }\textbf {\bibinfo {volume}
  {25}},\ \bibinfo {pages} {978--989} (\bibinfo {year} {1982})}\BibitemShut
  {NoStop}%
\bibitem [{\citenamefont {Zwanzig}(1983)}]{ZwanzigJCP1983}%
  \BibitemOpen
  \bibfield  {author} {\bibinfo {author} {\bibfnamefont {R.}~\bibnamefont
  {Zwanzig}},\ }\bibfield  {title} {\enquote {\bibinfo {title} {On the relation
  between self-diffusion and viscosity of liquids},}\ }\href {\doibase
  10.1063/1.446338} {\bibfield  {journal} {\bibinfo  {journal} {J. Chem.
  Phys.}\ }\textbf {\bibinfo {volume} {79}},\ \bibinfo {pages} {4507--4508}
  (\bibinfo {year} {1983})}\BibitemShut {NoStop}%
\bibitem [{\citenamefont {Khrapak}(2022{\natexlab{a}})}]{KhrapakApplSci2022}%
  \BibitemOpen
  \bibfield  {author} {\bibinfo {author} {\bibfnamefont {S.}~\bibnamefont
  {Khrapak}},\ }\bibfield  {title} {\enquote {\bibinfo {title} {Vibrational
  model of heat conduction in a fluid of hard spheres},}\ }\href {\doibase
  10.3390/app12157939} {\bibfield  {journal} {\bibinfo  {journal} {Appl. Sci.}\
  }\textbf {\bibinfo {volume} {12}},\ \bibinfo {pages} {7939} (\bibinfo {year}
  {2022}{\natexlab{a}})}\BibitemShut {NoStop}%
\bibitem [{\citenamefont {Costigliola}\ \emph {et~al.}(2019)\citenamefont
  {Costigliola}, \citenamefont {Heyes}, \citenamefont {Schr{\o}der},\ and\
  \citenamefont {Dyre}}]{CostigliolaJCP2019}%
  \BibitemOpen
  \bibfield  {author} {\bibinfo {author} {\bibfnamefont {L.}~\bibnamefont
  {Costigliola}}, \bibinfo {author} {\bibfnamefont {D.~M.}\ \bibnamefont
  {Heyes}}, \bibinfo {author} {\bibfnamefont {T.~B.}\ \bibnamefont
  {Schr{\o}der}}, \ and\ \bibinfo {author} {\bibfnamefont {J.~C.}\ \bibnamefont
  {Dyre}},\ }\bibfield  {title} {\enquote {\bibinfo {title} {Revisiting the
  {S}tokes-{E}instein relation without a hydrodynamic diameter},}\ }\href
  {\doibase 10.1063/1.5080662} {\bibfield  {journal} {\bibinfo  {journal} {J.
  Chem. Phys.}\ }\textbf {\bibinfo {volume} {150}},\ \bibinfo {pages} {021101}
  (\bibinfo {year} {2019})}\BibitemShut {NoStop}%
\bibitem [{\citenamefont {Khrapak}\ and\ \citenamefont
  {Khrapak}(2021{\natexlab{a}})}]{KhrapakPRE10_2021}%
  \BibitemOpen
  \bibfield  {author} {\bibinfo {author} {\bibfnamefont {S.~A.}\ \bibnamefont
  {Khrapak}}\ and\ \bibinfo {author} {\bibfnamefont {A.~G.}\ \bibnamefont
  {Khrapak}},\ }\bibfield  {title} {\enquote {\bibinfo {title} {Excess entropy
  and {S}tokes-{E}instein relation in simple fluids},}\ }\href {\doibase
  10.1103/physreve.104.044110} {\bibfield  {journal} {\bibinfo  {journal}
  {Phys. Rev. E}\ }\textbf {\bibinfo {volume} {104}},\ \bibinfo {pages}
  {044110} (\bibinfo {year} {2021}{\natexlab{a}})}\BibitemShut {NoStop}%
\bibitem [{\citenamefont
  {Khrapak}(2021{\natexlab{b}})}]{KhrapakMolecules12_2021}%
  \BibitemOpen
  \bibfield  {author} {\bibinfo {author} {\bibfnamefont {S.~A.}\ \bibnamefont
  {Khrapak}},\ }\bibfield  {title} {\enquote {\bibinfo {title} {Self-diffusion
  in simple liquids as a random walk process},}\ }\href {\doibase
  10.3390/molecules26247499} {\bibfield  {journal} {\bibinfo  {journal}
  {Molecules}\ }\textbf {\bibinfo {volume} {26}},\ \bibinfo {pages} {7499}
  (\bibinfo {year} {2021}{\natexlab{b}})}\BibitemShut {NoStop}%
\bibitem [{\citenamefont {Zaccone}\ and\ \citenamefont
  {Baggioli}(2021)}]{ZacconePNAS2021}%
  \BibitemOpen
  \bibfield  {author} {\bibinfo {author} {\bibfnamefont {A.}~\bibnamefont
  {Zaccone}}\ and\ \bibinfo {author} {\bibfnamefont {M.}~\bibnamefont
  {Baggioli}},\ }\bibfield  {title} {\enquote {\bibinfo {title} {Universal law
  for the vibrational density of states of liquids},}\ }\href {\doibase
  10.1073/pnas.2022303118} {\bibfield  {journal} {\bibinfo  {journal} {Proc.
  Natl. Acad. Sci.}\ }\textbf {\bibinfo {volume} {118}},\ \bibinfo {pages}
  {e2022303118} (\bibinfo {year} {2021})}\BibitemShut {NoStop}%
\bibitem [{\citenamefont {Stamper}\ \emph {et~al.}(2022)\citenamefont
  {Stamper}, \citenamefont {Cortie}, \citenamefont {Yue}, \citenamefont
  {Wang},\ and\ \citenamefont {Yu}}]{StamperJPCL2022}%
  \BibitemOpen
  \bibfield  {author} {\bibinfo {author} {\bibfnamefont {C.}~\bibnamefont
  {Stamper}}, \bibinfo {author} {\bibfnamefont {D.}~\bibnamefont {Cortie}},
  \bibinfo {author} {\bibfnamefont {Z.}~\bibnamefont {Yue}}, \bibinfo {author}
  {\bibfnamefont {X.}~\bibnamefont {Wang}}, \ and\ \bibinfo {author}
  {\bibfnamefont {D.}~\bibnamefont {Yu}},\ }\bibfield  {title} {\enquote
  {\bibinfo {title} {Experimental confirmation of the universal law for the
  vibrational density of states of liquids},}\ }\href {\doibase
  10.1021/acs.jpclett.2c00297} {\bibfield  {journal} {\bibinfo  {journal} {J.
  Phys. Chem. Lett.}\ }\textbf {\bibinfo {volume} {13}},\ \bibinfo {pages}
  {3105--3111} (\bibinfo {year} {2022})}\BibitemShut {NoStop}%
\bibitem [{\citenamefont {Rao}(1941)}]{Rao1941}%
  \BibitemOpen
  \bibfield  {author} {\bibinfo {author} {\bibfnamefont {M.~R.}\ \bibnamefont
  {Rao}},\ }\bibfield  {title} {\enquote {\bibinfo {title} {Thermal
  conductivity of liquids},}\ }\href {\doibase 10.1103/physrev.59.212}
  {\bibfield  {journal} {\bibinfo  {journal} {Phys. Rev.}\ }\textbf {\bibinfo
  {volume} {59}},\ \bibinfo {pages} {212--212} (\bibinfo {year}
  {1941})}\BibitemShut {NoStop}%
\bibitem [{\citenamefont {Horrocks}\ and\ \citenamefont
  {McLaughlin}(1960)}]{Horrocks1960}%
  \BibitemOpen
  \bibfield  {author} {\bibinfo {author} {\bibfnamefont {J.~K.}\ \bibnamefont
  {Horrocks}}\ and\ \bibinfo {author} {\bibfnamefont {E.}~\bibnamefont
  {McLaughlin}},\ }\bibfield  {title} {\enquote {\bibinfo {title} {Thermal
  conductivity of simple molecules in the condensed state},}\ }\href {\doibase
  10.1039/tf9605600206} {\bibfield  {journal} {\bibinfo  {journal} {Trans.
  Faraday Soc.}\ }\textbf {\bibinfo {volume} {56}},\ \bibinfo {pages} {206}
  (\bibinfo {year} {1960})}\BibitemShut {NoStop}%
\bibitem [{\citenamefont {Horrocks}\ and\ \citenamefont
  {McLaughlin}(1963)}]{HorrocksTFS1963}%
  \BibitemOpen
  \bibfield  {author} {\bibinfo {author} {\bibfnamefont {J.}~\bibnamefont
  {Horrocks}}\ and\ \bibinfo {author} {\bibfnamefont {E.}~\bibnamefont
  {McLaughlin}},\ }\bibfield  {title} {\enquote {\bibinfo {title} {Temperature
  dependence of the thermal conductivity of liquids},}\ }\href {\doibase
  10.1039/tf9635901709} {\bibfield  {journal} {\bibinfo  {journal} {Trans.
  Faraday Soc.}\ }\textbf {\bibinfo {volume} {59}},\ \bibinfo {pages} {1709}
  (\bibinfo {year} {1963})}\BibitemShut {NoStop}%
\bibitem [{\citenamefont {Khrapak}(2021{\natexlab{c}})}]{KhrapakPoP01_2021}%
  \BibitemOpen
  \bibfield  {author} {\bibinfo {author} {\bibfnamefont {S.~A.}\ \bibnamefont
  {Khrapak}},\ }\bibfield  {title} {\enquote {\bibinfo {title} {Thermal
  conduction in two-dimensional complex plasma layers},}\ }\href {\doibase
  10.1063/5.0038078} {\bibfield  {journal} {\bibinfo  {journal} {Phys.
  Plasmas}\ }\textbf {\bibinfo {volume} {28}},\ \bibinfo {pages} {010704}
  (\bibinfo {year} {2021}{\natexlab{c}})}\BibitemShut {NoStop}%
\bibitem [{\citenamefont {Khrapak}(2021{\natexlab{d}})}]{KhrapakPoP08_2021}%
  \BibitemOpen
  \bibfield  {author} {\bibinfo {author} {\bibfnamefont {S.~A.}\ \bibnamefont
  {Khrapak}},\ }\bibfield  {title} {\enquote {\bibinfo {title} {Thermal
  conductivity of strongly coupled {Y}ukawa fluids},}\ }\href {\doibase
  10.1063/5.0056763} {\bibfield  {journal} {\bibinfo  {journal} {Phys.
  Plasmas}\ }\textbf {\bibinfo {volume} {28}},\ \bibinfo {pages} {084501}
  (\bibinfo {year} {2021}{\natexlab{d}})}\BibitemShut {NoStop}%
\bibitem [{\citenamefont {Khrapak}(2023{\natexlab{a}})}]{KhrapakPPR2023}%
  \BibitemOpen
  \bibfield  {author} {\bibinfo {author} {\bibfnamefont {S.~A.}\ \bibnamefont
  {Khrapak}},\ }\bibfield  {title} {\enquote {\bibinfo {title} {Vibrational
  model of heat transfer in strongly coupled {Y}ukawa fluids (dusty plasma
  liquids)},}\ }\href {\doibase 10.1134/s1063780x22600876} {\bibfield
  {journal} {\bibinfo  {journal} {Plasma Phys. Rep.}\ }\textbf {\bibinfo
  {volume} {49}},\ \bibinfo {pages} {15--22} (\bibinfo {year}
  {2023}{\natexlab{a}})}\BibitemShut {NoStop}%
\bibitem [{\citenamefont {da~C.~Andrade}(1952)}]{Andrade1952}%
  \BibitemOpen
  \bibfield  {author} {\bibinfo {author} {\bibfnamefont {E.~N.}\ \bibnamefont
  {da~C.~Andrade}},\ }\bibfield  {title} {\enquote {\bibinfo {title} {Viscosity
  and thermal conductivity of liquid argon},}\ }\href {\doibase
  10.1038/170794b0} {\bibfield  {journal} {\bibinfo  {journal} {Nature}\
  }\textbf {\bibinfo {volume} {170}},\ \bibinfo {pages} {794--794} (\bibinfo
  {year} {1952})}\BibitemShut {NoStop}%
\bibitem [{\citenamefont {Khrapak}\ and\ \citenamefont
  {Khrapak}(2022{\natexlab{a}})}]{KhrapakPoF2022}%
  \BibitemOpen
  \bibfield  {author} {\bibinfo {author} {\bibfnamefont {S.~A.}\ \bibnamefont
  {Khrapak}}\ and\ \bibinfo {author} {\bibfnamefont {A.~G.}\ \bibnamefont
  {Khrapak}},\ }\bibfield  {title} {\enquote {\bibinfo {title} {Minima of shear
  viscosity and thermal conductivity coefficients of classical fluids},}\
  }\href {\doibase 10.1063/5.0082465} {\bibfield  {journal} {\bibinfo
  {journal} {Phys. Fluids}\ }\textbf {\bibinfo {volume} {34}},\ \bibinfo
  {pages} {027102} (\bibinfo {year} {2022}{\natexlab{a}})}\BibitemShut
  {NoStop}%
\bibitem [{\citenamefont {Khrapak}\ and\ \citenamefont
  {Khrapak}(2022{\natexlab{b}})}]{KhrapakJCP2022_1}%
  \BibitemOpen
  \bibfield  {author} {\bibinfo {author} {\bibfnamefont {S.~A.}\ \bibnamefont
  {Khrapak}}\ and\ \bibinfo {author} {\bibfnamefont {A.~G.}\ \bibnamefont
  {Khrapak}},\ }\bibfield  {title} {\enquote {\bibinfo {title} {Freezing
  density scaling of fluid transport properties: Application to liquefied noble
  gases},}\ }\href {\doibase 10.1063/5.0096947} {\bibfield  {journal} {\bibinfo
   {journal} {J. Chem. Phys.}\ }\textbf {\bibinfo {volume} {157}},\ \bibinfo
  {pages} {014501} (\bibinfo {year} {2022}{\natexlab{b}})}\BibitemShut
  {NoStop}%
\bibitem [{\citenamefont {Khrapak}(2023{\natexlab{b}})}]{KhrapakJMolLiq2023}%
  \BibitemOpen
  \bibfield  {author} {\bibinfo {author} {\bibfnamefont {S.A.}\ \bibnamefont
  {Khrapak}},\ }\bibfield  {title} {\enquote {\bibinfo {title} {Bridgman
  formula for the thermal conductivity of atomic and molecular liquids},}\
  }\href {\doibase 10.1016/j.molliq.2023.121786} {\bibfield  {journal}
  {\bibinfo  {journal} {J. Mol. Liq.}\ }\textbf {\bibinfo {volume} {381}},\
  \bibinfo {pages} {121786} (\bibinfo {year} {2023}{\natexlab{b}})}\BibitemShut
  {NoStop}%
\bibitem [{\citenamefont {Bridgman}(1923)}]{Bridgman1923}%
  \BibitemOpen
  \bibfield  {author} {\bibinfo {author} {\bibfnamefont {P.~W.}\ \bibnamefont
  {Bridgman}},\ }\bibfield  {title} {\enquote {\bibinfo {title} {The thermal
  conductivity of liquids under pressure},}\ }\href {\doibase 10.2307/20026073}
  {\bibfield  {journal} {\bibinfo  {journal} {PNAAS}\ }\textbf {\bibinfo
  {volume} {59}},\ \bibinfo {pages} {141} (\bibinfo {year} {1923})}\BibitemShut
  {NoStop}%
\bibitem [{\citenamefont {Zhao}\ \emph {et~al.}(2021)\citenamefont {Zhao},
  \citenamefont {Wingert}, \citenamefont {Chen},\ and\ \citenamefont
  {Garay}}]{ZhaoJAP2021}%
  \BibitemOpen
  \bibfield  {author} {\bibinfo {author} {\bibfnamefont {A.~Z.}\ \bibnamefont
  {Zhao}}, \bibinfo {author} {\bibfnamefont {M.~C.}\ \bibnamefont {Wingert}},
  \bibinfo {author} {\bibfnamefont {R.}~\bibnamefont {Chen}}, \ and\ \bibinfo
  {author} {\bibfnamefont {J.~E.}\ \bibnamefont {Garay}},\ }\bibfield  {title}
  {\enquote {\bibinfo {title} {Phonon gas model for thermal conductivity of
  dense, strongly interacting liquids},}\ }\href {\doibase 10.1063/5.0040734}
  {\bibfield  {journal} {\bibinfo  {journal} {J. Appl. Phys.}\ }\textbf
  {\bibinfo {volume} {129}},\ \bibinfo {pages} {235101} (\bibinfo {year}
  {2021})}\BibitemShut {NoStop}%
\bibitem [{\citenamefont {Xi}\ \emph {et~al.}(2020)\citenamefont {Xi},
  \citenamefont {Zhong}, \citenamefont {He}, \citenamefont {Xu}, \citenamefont
  {Nakayama}, \citenamefont {Wang}, \citenamefont {Liu}, \citenamefont {Zhou},\
  and\ \citenamefont {Li}}]{XiCPL2020}%
  \BibitemOpen
  \bibfield  {author} {\bibinfo {author} {\bibfnamefont {Q.}~\bibnamefont
  {Xi}}, \bibinfo {author} {\bibfnamefont {J.}~\bibnamefont {Zhong}}, \bibinfo
  {author} {\bibfnamefont {J.}~\bibnamefont {He}}, \bibinfo {author}
  {\bibfnamefont {X.}~\bibnamefont {Xu}}, \bibinfo {author} {\bibfnamefont
  {T.}~\bibnamefont {Nakayama}}, \bibinfo {author} {\bibfnamefont
  {Y.}~\bibnamefont {Wang}}, \bibinfo {author} {\bibfnamefont {J.}~\bibnamefont
  {Liu}}, \bibinfo {author} {\bibfnamefont {J.}~\bibnamefont {Zhou}}, \ and\
  \bibinfo {author} {\bibfnamefont {B.}~\bibnamefont {Li}},\ }\bibfield
  {title} {\enquote {\bibinfo {title} {A ubiquitous thermal conductivity
  formula for liquids, polymer glass, and amorphous solids},}\ }\href {\doibase
  10.1088/0256-307x/37/10/104401} {\bibfield  {journal} {\bibinfo  {journal}
  {Chin. Phys. Lett.}\ }\textbf {\bibinfo {volume} {37}},\ \bibinfo {pages}
  {104401} (\bibinfo {year} {2020})}\BibitemShut {NoStop}%
\bibitem [{\citenamefont {Bird}\ \emph {et~al.}(2002)\citenamefont {Bird},
  \citenamefont {Lightfoot},\ and\ \citenamefont {Stewart}}]{BirdBook}%
  \BibitemOpen
  \bibfield  {author} {\bibinfo {author} {\bibfnamefont {R.~B.}\ \bibnamefont
  {Bird}}, \bibinfo {author} {\bibfnamefont {E.~N.}\ \bibnamefont {Lightfoot}},
  \ and\ \bibinfo {author} {\bibfnamefont {W.~E.}\ \bibnamefont {Stewart}},\
  }\href@noop {} {\emph {\bibinfo {title} {Transport Phenomena -}}}\ (\bibinfo
  {publisher} {J. Wiley},\ \bibinfo {address} {New York},\ \bibinfo {year}
  {2002})\BibitemShut {NoStop}%
\bibitem [{\citenamefont {Khrapak}\ and\ \citenamefont
  {Khrapak}(2023)}]{KhrapakPoF2023}%
  \BibitemOpen
  \bibfield  {author} {\bibinfo {author} {\bibfnamefont {S.~A.}\ \bibnamefont
  {Khrapak}}\ and\ \bibinfo {author} {\bibfnamefont {A.~G.}\ \bibnamefont
  {Khrapak}},\ }\bibfield  {title} {\enquote {\bibinfo {title} {Sound
  velocities in liquids near freezing: {D}ependence on the interaction
  potential and correlations with thermal conductivity},}\ }\href {\doibase
  10.1063/5.0157945} {\bibfield  {journal} {\bibinfo  {journal} {Phys. Fluids}\
  }\textbf {\bibinfo {volume} {35}},\ \bibinfo {pages} {077129} (\bibinfo
  {year} {2023})}\BibitemShut {NoStop}%
\bibitem [{\citenamefont {Hansen}\ and\ \citenamefont
  {McDonald}(2006)}]{HansenBook}%
  \BibitemOpen
  \bibfield  {author} {\bibinfo {author} {\bibfnamefont {J.-P.}\ \bibnamefont
  {Hansen}}\ and\ \bibinfo {author} {\bibfnamefont {I.~R.}\ \bibnamefont
  {McDonald}},\ }\href@noop {} {\emph {\bibinfo {title} {Theory of Simple
  Liquids -}}}\ (\bibinfo  {publisher} {Elsevier},\ \bibinfo {address}
  {Amsterdam},\ \bibinfo {year} {2006})\BibitemShut {NoStop}%
\bibitem [{\citenamefont {Balucani}\ and\ \citenamefont
  {Zoppi}(1994)}]{BalucaniBook}%
  \BibitemOpen
  \bibfield  {author} {\bibinfo {author} {\bibfnamefont {U.}~\bibnamefont
  {Balucani}}\ and\ \bibinfo {author} {\bibfnamefont {M.}~\bibnamefont
  {Zoppi}},\ }\href@noop {} {\emph {\bibinfo {title} {Dynamics of the Liquid
  State}}}\ (\bibinfo  {publisher} {Clarendon Press},\ \bibinfo {address}
  {Oxford},\ \bibinfo {year} {1994})\BibitemShut {NoStop}%
\bibitem [{\citenamefont {Ohta}\ and\ \citenamefont
  {Hamaguchi}(2000)}]{OhtaPRL2000}%
  \BibitemOpen
  \bibfield  {author} {\bibinfo {author} {\bibfnamefont {H.}~\bibnamefont
  {Ohta}}\ and\ \bibinfo {author} {\bibfnamefont {S.}~\bibnamefont
  {Hamaguchi}},\ }\bibfield  {title} {\enquote {\bibinfo {title} {Wave
  dispersion relations in {Y}ukawa fluids},}\ }\href {\doibase
  10.1103/physrevlett.84.6026} {\bibfield  {journal} {\bibinfo  {journal}
  {Phys. Rev. Lett.}\ }\textbf {\bibinfo {volume} {84}},\ \bibinfo {pages}
  {6026--6029} (\bibinfo {year} {2000})}\BibitemShut {NoStop}%
\bibitem [{\citenamefont {Hosokawa}\ \emph {et~al.}(2015)\citenamefont
  {Hosokawa}, \citenamefont {Inui}, \citenamefont {Kajihara}, \citenamefont
  {Tsutsui},\ and\ \citenamefont {Baron}}]{HosokawaJPCM2015}%
  \BibitemOpen
  \bibfield  {author} {\bibinfo {author} {\bibfnamefont {S.}~\bibnamefont
  {Hosokawa}}, \bibinfo {author} {\bibfnamefont {M.}~\bibnamefont {Inui}},
  \bibinfo {author} {\bibfnamefont {Y.}~\bibnamefont {Kajihara}}, \bibinfo
  {author} {\bibfnamefont {S.}~\bibnamefont {Tsutsui}}, \ and\ \bibinfo
  {author} {\bibfnamefont {A.~Q.~R.}\ \bibnamefont {Baron}},\ }\bibfield
  {title} {\enquote {\bibinfo {title} {Transverse excitations in liquid {Fe},
  {Cu} and {Zn}},}\ }\href {\doibase 10.1088/0953-8984/27/19/194104} {\bibfield
   {journal} {\bibinfo  {journal} {J. Phys.: Condens. Matter}\ }\textbf
  {\bibinfo {volume} {27}},\ \bibinfo {pages} {194104} (\bibinfo {year}
  {2015})}\BibitemShut {NoStop}%
\bibitem [{\citenamefont {Bryk}\ \emph {et~al.}(2017)\citenamefont {Bryk},
  \citenamefont {Huerta}, \citenamefont {Hordiichuk},\ and\ \citenamefont
  {Trokhymchuk}}]{BrykJCP2017}%
  \BibitemOpen
  \bibfield  {author} {\bibinfo {author} {\bibfnamefont {T.}~\bibnamefont
  {Bryk}}, \bibinfo {author} {\bibfnamefont {A.}~\bibnamefont {Huerta}},
  \bibinfo {author} {\bibfnamefont {V.}~\bibnamefont {Hordiichuk}}, \ and\
  \bibinfo {author} {\bibfnamefont {A.~D.}\ \bibnamefont {Trokhymchuk}},\
  }\bibfield  {title} {\enquote {\bibinfo {title} {Non-hydrodynamic transverse
  collective excitations in hard-sphere fluids},}\ }\href {\doibase
  10.1063/1.4997640} {\bibfield  {journal} {\bibinfo  {journal} {J. Chem.
  Phys.}\ }\textbf {\bibinfo {volume} {147}},\ \bibinfo {pages} {064509}
  (\bibinfo {year} {2017})}\BibitemShut {NoStop}%
\bibitem [{\citenamefont {Bolmatov}\ \emph {et~al.}(2012)\citenamefont
  {Bolmatov}, \citenamefont {Brazhkin},\ and\ \citenamefont
  {Trachenko}}]{BolmatovSciRep2012}%
  \BibitemOpen
  \bibfield  {author} {\bibinfo {author} {\bibfnamefont {D.}~\bibnamefont
  {Bolmatov}}, \bibinfo {author} {\bibfnamefont {V.~V.}\ \bibnamefont
  {Brazhkin}}, \ and\ \bibinfo {author} {\bibfnamefont {K.}~\bibnamefont
  {Trachenko}},\ }\bibfield  {title} {\enquote {\bibinfo {title} {The phonon
  theory of liquid thermodynamics},}\ }\href {\doibase 10.1038/srep00421}
  {\bibfield  {journal} {\bibinfo  {journal} {Sci. Rep.}\ }\textbf {\bibinfo
  {volume} {2}},\ \bibinfo {pages} {421} (\bibinfo {year} {2012})}\BibitemShut
  {NoStop}%
\bibitem [{\citenamefont {Khrapak}\ and\ \citenamefont
  {Yurchenko}(2021)}]{KhrapakJCP2021}%
  \BibitemOpen
  \bibfield  {author} {\bibinfo {author} {\bibfnamefont {S.~A.}\ \bibnamefont
  {Khrapak}}\ and\ \bibinfo {author} {\bibfnamefont {S.~O.}\ \bibnamefont
  {Yurchenko}},\ }\bibfield  {title} {\enquote {\bibinfo {title} {Entropy of
  simple fluids with repulsive interactions near freezing},}\ }\href {\doibase
  10.1063/5.0063559} {\bibfield  {journal} {\bibinfo  {journal} {J. Chem.
  Phys.}\ }\textbf {\bibinfo {volume} {155}},\ \bibinfo {pages} {134501}
  (\bibinfo {year} {2021})}\BibitemShut {NoStop}%
\bibitem [{\citenamefont {Cahill}\ and\ \citenamefont
  {Pohl}(1989)}]{Cahill1989}%
  \BibitemOpen
  \bibfield  {author} {\bibinfo {author} {\bibfnamefont {D.~G.}\ \bibnamefont
  {Cahill}}\ and\ \bibinfo {author} {\bibfnamefont {R.O.}\ \bibnamefont
  {Pohl}},\ }\bibfield  {title} {\enquote {\bibinfo {title} {Heat flow and
  lattice vibrations in glasses},}\ }\href {\doibase
  10.1016/0038-1098(89)90630-3} {\bibfield  {journal} {\bibinfo  {journal}
  {Solid State Commun.}\ }\textbf {\bibinfo {volume} {70}},\ \bibinfo {pages}
  {927--930} (\bibinfo {year} {1989})}\BibitemShut {NoStop}%
\bibitem [{\citenamefont {Cahill}\ \emph {et~al.}(1992)\citenamefont {Cahill},
  \citenamefont {Watson},\ and\ \citenamefont {Pohl}}]{Cahill1992}%
  \BibitemOpen
  \bibfield  {author} {\bibinfo {author} {\bibfnamefont {D.~G.}\ \bibnamefont
  {Cahill}}, \bibinfo {author} {\bibfnamefont {S.~K.}\ \bibnamefont {Watson}},
  \ and\ \bibinfo {author} {\bibfnamefont {R.~O.}\ \bibnamefont {Pohl}},\
  }\bibfield  {title} {\enquote {\bibinfo {title} {Lower limit to the thermal
  conductivity of disordered crystals},}\ }\href {\doibase
  10.1103/physrevb.46.6131} {\bibfield  {journal} {\bibinfo  {journal} {Phys.
  Rev. B}\ }\textbf {\bibinfo {volume} {46}},\ \bibinfo {pages} {6131--6140}
  (\bibinfo {year} {1992})}\BibitemShut {NoStop}%
\bibitem [{\citenamefont {Goree}\ \emph {et~al.}(2012)\citenamefont {Goree},
  \citenamefont {Donk{\'{o}}},\ and\ \citenamefont {Hartmann}}]{GoreePRE2012}%
  \BibitemOpen
  \bibfield  {author} {\bibinfo {author} {\bibfnamefont {J.}~\bibnamefont
  {Goree}}, \bibinfo {author} {\bibfnamefont {Z.}~\bibnamefont {Donk{\'{o}}}},
  \ and\ \bibinfo {author} {\bibfnamefont {P.}~\bibnamefont {Hartmann}},\
  }\bibfield  {title} {\enquote {\bibinfo {title} {Cutoff wave number for shear
  waves and {M}axwell relaxation time in {Y}ukawa liquids},}\ }\href {\doibase
  10.1103/physreve.85.066401} {\bibfield  {journal} {\bibinfo  {journal} {Phys.
  Rev. E}\ }\textbf {\bibinfo {volume} {85}},\ \bibinfo {pages} {066401}
  (\bibinfo {year} {2012})}\BibitemShut {NoStop}%
\bibitem [{\citenamefont {Yang}\ \emph {et~al.}(2017)\citenamefont {Yang},
  \citenamefont {Dove}, \citenamefont {Brazhkin},\ and\ \citenamefont
  {Trachenko}}]{YangPRL2017}%
  \BibitemOpen
  \bibfield  {author} {\bibinfo {author} {\bibfnamefont {C.}~\bibnamefont
  {Yang}}, \bibinfo {author} {\bibfnamefont {M.~T.}\ \bibnamefont {Dove}},
  \bibinfo {author} {\bibfnamefont {V.~V.}\ \bibnamefont {Brazhkin}}, \ and\
  \bibinfo {author} {\bibfnamefont {K.}~\bibnamefont {Trachenko}},\ }\bibfield
  {title} {\enquote {\bibinfo {title} {Emergence and evolution of the k-gap in
  spectra of liquid and supercritical states},}\ }\href {\doibase
  10.1103/physrevlett.118.215502} {\bibfield  {journal} {\bibinfo  {journal}
  {Phys. Rev. Lett.}\ }\textbf {\bibinfo {volume} {118}},\ \bibinfo {pages}
  {215502} (\bibinfo {year} {2017})}\BibitemShut {NoStop}%
\bibitem [{\citenamefont {Trachenko}\ and\ \citenamefont
  {Brazhkin}(2015)}]{TrachenkoRPP2015}%
  \BibitemOpen
  \bibfield  {author} {\bibinfo {author} {\bibfnamefont {K.}~\bibnamefont
  {Trachenko}}\ and\ \bibinfo {author} {\bibfnamefont {V.~V.}\ \bibnamefont
  {Brazhkin}},\ }\bibfield  {title} {\enquote {\bibinfo {title} {Collective
  modes and thermodynamics of the liquid state},}\ }\href {\doibase
  10.1088/0034-4885/79/1/016502} {\bibfield  {journal} {\bibinfo  {journal}
  {Rep. Progr. Phys.}\ }\textbf {\bibinfo {volume} {79}},\ \bibinfo {pages}
  {016502} (\bibinfo {year} {2015})}\BibitemShut {NoStop}%
\bibitem [{\citenamefont {Khrapak}\ \emph {et~al.}(2019)\citenamefont
  {Khrapak}, \citenamefont {Khrapak}, \citenamefont {Kryuchkov},\ and\
  \citenamefont {Yurchenko}}]{KhrapakJCP2019}%
  \BibitemOpen
  \bibfield  {author} {\bibinfo {author} {\bibfnamefont {S.~A.}\ \bibnamefont
  {Khrapak}}, \bibinfo {author} {\bibfnamefont {A.~G.}\ \bibnamefont
  {Khrapak}}, \bibinfo {author} {\bibfnamefont {N.~P.}\ \bibnamefont
  {Kryuchkov}}, \ and\ \bibinfo {author} {\bibfnamefont {S.~O.}\ \bibnamefont
  {Yurchenko}},\ }\bibfield  {title} {\enquote {\bibinfo {title} {Onset of
  transverse (shear) waves in strongly-coupled {Y}ukawa fluids},}\ }\href
  {\doibase 10.1063/1.5088141} {\bibfield  {journal} {\bibinfo  {journal} {J.
  Chem. Phys.}\ }\textbf {\bibinfo {volume} {150}},\ \bibinfo {pages} {104503}
  (\bibinfo {year} {2019})}\BibitemShut {NoStop}%
\bibitem [{\citenamefont {Kryuchkov}\ \emph {et~al.}(2019)\citenamefont
  {Kryuchkov}, \citenamefont {Mistryukova}, \citenamefont {Brazhkin},\ and\
  \citenamefont {Yurchenko}}]{KryuchkovSciRep2019}%
  \BibitemOpen
  \bibfield  {author} {\bibinfo {author} {\bibfnamefont {N.~P.}\ \bibnamefont
  {Kryuchkov}}, \bibinfo {author} {\bibfnamefont {L.~A.}\ \bibnamefont
  {Mistryukova}}, \bibinfo {author} {\bibfnamefont {V.~V.}\ \bibnamefont
  {Brazhkin}}, \ and\ \bibinfo {author} {\bibfnamefont {S.~O.}\ \bibnamefont
  {Yurchenko}},\ }\bibfield  {title} {\enquote {\bibinfo {title} {Excitation
  spectra in fluids: How to analyze them properly},}\ }\href {\doibase
  10.1038/s41598-019-46979-y} {\bibfield  {journal} {\bibinfo  {journal} {Sci.
  Rep.}\ }\textbf {\bibinfo {volume} {9}},\ \bibinfo {pages} {10483} (\bibinfo
  {year} {2019})}\BibitemShut {NoStop}%
\bibitem [{\citenamefont {Kryuchkov}\ and\ \citenamefont
  {Yurchenko}(2021)}]{KryuchkovJCP2021}%
  \BibitemOpen
  \bibfield  {author} {\bibinfo {author} {\bibfnamefont {N.~P.}\ \bibnamefont
  {Kryuchkov}}\ and\ \bibinfo {author} {\bibfnamefont {S.~O.}\ \bibnamefont
  {Yurchenko}},\ }\bibfield  {title} {\enquote {\bibinfo {title} {Collective
  excitations in active fluids: Microflows and breakdown in spectral
  equipartition of kinetic energy},}\ }\href {\doibase 10.1063/5.0054854}
  {\bibfield  {journal} {\bibinfo  {journal} {J. Chem. Phys.}\ }\textbf
  {\bibinfo {volume} {155}},\ \bibinfo {pages} {024902} (\bibinfo {year}
  {2021})}\BibitemShut {NoStop}%
\bibitem [{\citenamefont {Khrapak}\ and\ \citenamefont
  {Khrapak}(2021{\natexlab{b}})}]{KhrapakPRE04_2021}%
  \BibitemOpen
  \bibfield  {author} {\bibinfo {author} {\bibfnamefont {S.~A.}\ \bibnamefont
  {Khrapak}}\ and\ \bibinfo {author} {\bibfnamefont {A.~G.}\ \bibnamefont
  {Khrapak}},\ }\bibfield  {title} {\enquote {\bibinfo {title} {Transport
  properties of {L}ennard-{J}ones fluids: Freezing density scaling along
  isotherms},}\ }\href {\doibase 10.1103/physreve.103.042122} {\bibfield
  {journal} {\bibinfo  {journal} {Phys. Rev. E}\ }\textbf {\bibinfo {volume}
  {103}},\ \bibinfo {pages} {042122} (\bibinfo {year}
  {2021}{\natexlab{b}})}\BibitemShut {NoStop}%
\bibitem [{\citenamefont {Khrapak}\ and\ \citenamefont
  {Khrapak}(2022{\natexlab{c}})}]{KhrapakJPCL2022}%
  \BibitemOpen
  \bibfield  {author} {\bibinfo {author} {\bibfnamefont {S.~A.}\ \bibnamefont
  {Khrapak}}\ and\ \bibinfo {author} {\bibfnamefont {A.~G.}\ \bibnamefont
  {Khrapak}},\ }\bibfield  {title} {\enquote {\bibinfo {title} {Freezing
  temperature and density scaling of transport coefficients},}\ }\href
  {\doibase 10.1021/acs.jpclett.2c00408} {\bibfield  {journal} {\bibinfo
  {journal} {J. Phys. Chem. Lett.}\ }\textbf {\bibinfo {volume} {13}},\
  \bibinfo {pages} {2674--2678} (\bibinfo {year}
  {2022}{\natexlab{c}})}\BibitemShut {NoStop}%
\bibitem [{\citenamefont {Bell}\ \emph {et~al.}(2019)\citenamefont {Bell},
  \citenamefont {Messerly}, \citenamefont {Thol}, \citenamefont {Costigliola},\
  and\ \citenamefont {Dyre}}]{BellJPCB2019}%
  \BibitemOpen
  \bibfield  {author} {\bibinfo {author} {\bibfnamefont {I.~H.}\ \bibnamefont
  {Bell}}, \bibinfo {author} {\bibfnamefont {R.}~\bibnamefont {Messerly}},
  \bibinfo {author} {\bibfnamefont {M.}~\bibnamefont {Thol}}, \bibinfo {author}
  {\bibfnamefont {L.}~\bibnamefont {Costigliola}}, \ and\ \bibinfo {author}
  {\bibfnamefont {J.~C.}\ \bibnamefont {Dyre}},\ }\bibfield  {title} {\enquote
  {\bibinfo {title} {Modified entropy scaling of the transport properties of
  the {L}ennard-{J}ones fluid},}\ }\href {\doibase 10.1021/acs.jpcb.9b05808}
  {\bibfield  {journal} {\bibinfo  {journal} {J. Phys. Chem. B}\ }\textbf
  {\bibinfo {volume} {123}},\ \bibinfo {pages} {6345--6363} (\bibinfo {year}
  {2019})}\BibitemShut {NoStop}%
\bibitem [{\citenamefont {Harris}(2020)}]{HarrisJCP2020}%
  \BibitemOpen
  \bibfield  {author} {\bibinfo {author} {\bibfnamefont {K.~R.}\ \bibnamefont
  {Harris}},\ }\bibfield  {title} {\enquote {\bibinfo {title} {Thermodynamic or
  density scaling of the thermal conductivity of liquids},}\ }\href {\doibase
  10.1063/5.0016389} {\bibfield  {journal} {\bibinfo  {journal} {J. Chem.
  Phys.}\ }\textbf {\bibinfo {volume} {153}},\ \bibinfo {pages} {104504}
  (\bibinfo {year} {2020})}\BibitemShut {NoStop}%
\bibitem [{\citenamefont {Allers}\ \emph {et~al.}(2020)\citenamefont {Allers},
  \citenamefont {Harvey}, \citenamefont {Garzon},\ and\ \citenamefont
  {Alam}}]{AllersJCP2020}%
  \BibitemOpen
  \bibfield  {author} {\bibinfo {author} {\bibfnamefont {J.~P.}\ \bibnamefont
  {Allers}}, \bibinfo {author} {\bibfnamefont {J.~A.}\ \bibnamefont {Harvey}},
  \bibinfo {author} {\bibfnamefont {F.~H.}\ \bibnamefont {Garzon}}, \ and\
  \bibinfo {author} {\bibfnamefont {T.~M.}\ \bibnamefont {Alam}},\ }\bibfield
  {title} {\enquote {\bibinfo {title} {Machine learning prediction of
  self-diffusion in {L}ennard-{J}ones fluids},}\ }\href {\doibase
  10.1063/5.0011512} {\bibfield  {journal} {\bibinfo  {journal} {J. Chem.
  Phys.}\ }\textbf {\bibinfo {volume} {153}},\ \bibinfo {pages} {034102}
  (\bibinfo {year} {2020})}\BibitemShut {NoStop}%
\bibitem [{\citenamefont {Meier}(2002)}]{Meier2002}%
  \BibitemOpen
  \bibfield  {author} {\bibinfo {author} {\bibfnamefont {K.}~\bibnamefont
  {Meier}},\ }\href@noop {} {\emph {\bibinfo {title} {Computer Simulation and
  Interpretation of the Transport Coefficients of the {L}ennard-{J}ones Model
  Fluid (PhD Thesis)}}}\ (\bibinfo  {publisher} {Shaker},\ \bibinfo {address}
  {Aachen},\ \bibinfo {year} {2002})\BibitemShut {NoStop}%
\bibitem [{\citenamefont {Meier}\ \emph
  {et~al.}(2004{\natexlab{a}})\citenamefont {Meier}, \citenamefont {Laesecke},\
  and\ \citenamefont {Kabelac}}]{MeierJCP_1}%
  \BibitemOpen
  \bibfield  {author} {\bibinfo {author} {\bibfnamefont {K.}~\bibnamefont
  {Meier}}, \bibinfo {author} {\bibfnamefont {A.}~\bibnamefont {Laesecke}}, \
  and\ \bibinfo {author} {\bibfnamefont {S.}~\bibnamefont {Kabelac}},\
  }\bibfield  {title} {\enquote {\bibinfo {title} {Transport coefficients of
  the {L}ennard-{J}ones model fluid. {I}. {V}iscosity},}\ }\href {\doibase
  10.1063/1.1770695} {\bibfield  {journal} {\bibinfo  {journal} {J. Chem.
  Phys.}\ }\textbf {\bibinfo {volume} {121}},\ \bibinfo {pages} {3671--3687}
  (\bibinfo {year} {2004}{\natexlab{a}})}\BibitemShut {NoStop}%
\bibitem [{\citenamefont {Meier}\ \emph
  {et~al.}(2004{\natexlab{b}})\citenamefont {Meier}, \citenamefont {Laesecke},\
  and\ \citenamefont {Kabelac}}]{MeierJCP_2}%
  \BibitemOpen
  \bibfield  {author} {\bibinfo {author} {\bibfnamefont {K.}~\bibnamefont
  {Meier}}, \bibinfo {author} {\bibfnamefont {A.}~\bibnamefont {Laesecke}}, \
  and\ \bibinfo {author} {\bibfnamefont {S.}~\bibnamefont {Kabelac}},\
  }\bibfield  {title} {\enquote {\bibinfo {title} {Transport coefficients of
  the {L}ennard-{J}ones model fluid. {II}. {S}elf-diffusion},}\ }\href
  {\doibase 10.1063/1.1786579} {\bibfield  {journal} {\bibinfo  {journal} {J.
  Chem. Phys.}\ }\textbf {\bibinfo {volume} {121}},\ \bibinfo {pages}
  {9526--9535} (\bibinfo {year} {2004}{\natexlab{b}})}\BibitemShut {NoStop}%
\bibitem [{\citenamefont {Baidakov}\ \emph {et~al.}(2011)\citenamefont
  {Baidakov}, \citenamefont {Protsenko},\ and\ \citenamefont
  {Kozlova}}]{BaidakovFPE2011}%
  \BibitemOpen
  \bibfield  {author} {\bibinfo {author} {\bibfnamefont {V.G.}\ \bibnamefont
  {Baidakov}}, \bibinfo {author} {\bibfnamefont {S.P.}\ \bibnamefont
  {Protsenko}}, \ and\ \bibinfo {author} {\bibfnamefont {Z.R.}\ \bibnamefont
  {Kozlova}},\ }\bibfield  {title} {\enquote {\bibinfo {title} {The
  self-diffusion coefficient in stable and metastable states of the
  {L}ennard{\textendash}{J}ones fluid},}\ }\href {\doibase
  10.1016/j.fluid.2011.03.002} {\bibfield  {journal} {\bibinfo  {journal}
  {Fluid Phase Equilibria}\ }\textbf {\bibinfo {volume} {305}},\ \bibinfo
  {pages} {106--113} (\bibinfo {year} {2011})}\BibitemShut {NoStop}%
\bibitem [{\citenamefont {Baidakov}\ \emph {et~al.}(2012)\citenamefont
  {Baidakov}, \citenamefont {Protsenko},\ and\ \citenamefont
  {Kozlova}}]{BaidakovJCP2012}%
  \BibitemOpen
  \bibfield  {author} {\bibinfo {author} {\bibfnamefont {V.~G.}\ \bibnamefont
  {Baidakov}}, \bibinfo {author} {\bibfnamefont {S.~P.}\ \bibnamefont
  {Protsenko}}, \ and\ \bibinfo {author} {\bibfnamefont {Z.~R.}\ \bibnamefont
  {Kozlova}},\ }\bibfield  {title} {\enquote {\bibinfo {title} {Metastable
  {L}ennard-{J}ones fluids. {I}. {S}hear viscosity},}\ }\href {\doibase
  10.1063/1.4758806} {\bibfield  {journal} {\bibinfo  {journal} {J. Chem.
  Phys.}\ }\textbf {\bibinfo {volume} {137}},\ \bibinfo {pages} {164507}
  (\bibinfo {year} {2012})}\BibitemShut {NoStop}%
\bibitem [{\citenamefont {Baidakov}\ and\ \citenamefont
  {Protsenko}(2014)}]{BaidakovJCP2014}%
  \BibitemOpen
  \bibfield  {author} {\bibinfo {author} {\bibfnamefont {V.~G.}\ \bibnamefont
  {Baidakov}}\ and\ \bibinfo {author} {\bibfnamefont {S.~P.}\ \bibnamefont
  {Protsenko}},\ }\bibfield  {title} {\enquote {\bibinfo {title} {Metastable
  {L}ennard-{J}ones fluids. {II}. {T}hermal conductivity},}\ }\href {\doibase
  10.1063/1.4880958} {\bibfield  {journal} {\bibinfo  {journal} {J. Chem.
  Phys.}\ }\textbf {\bibinfo {volume} {140}},\ \bibinfo {pages} {214506}
  (\bibinfo {year} {2014})}\BibitemShut {NoStop}%
\bibitem [{\citenamefont {Sousa}\ \emph {et~al.}(2012)\citenamefont {Sousa},
  \citenamefont {Ferreira},\ and\ \citenamefont {Barroso}}]{SousaJCP2012}%
  \BibitemOpen
  \bibfield  {author} {\bibinfo {author} {\bibfnamefont {J.~M.~G.}\
  \bibnamefont {Sousa}}, \bibinfo {author} {\bibfnamefont {A.~L.}\ \bibnamefont
  {Ferreira}}, \ and\ \bibinfo {author} {\bibfnamefont {M.~A.}\ \bibnamefont
  {Barroso}},\ }\bibfield  {title} {\enquote {\bibinfo {title} {Determination
  of the solid-fluid coexistence of the n - 6 {L}ennard-{J}ones system from
  free energy calculations},}\ }\href {\doibase 10.1063/1.4707746} {\bibfield
  {journal} {\bibinfo  {journal} {J. Chem. Phys.}\ }\textbf {\bibinfo {volume}
  {136}},\ \bibinfo {pages} {174502} (\bibinfo {year} {2012})}\BibitemShut
  {NoStop}%
\bibitem [{\citenamefont {Heyes}\ \emph {et~al.}(2019)\citenamefont {Heyes},
  \citenamefont {Dini}, \citenamefont {Costigliola},\ and\ \citenamefont
  {Dyre}}]{HeyesJCP2019}%
  \BibitemOpen
  \bibfield  {author} {\bibinfo {author} {\bibfnamefont {D.~M.}\ \bibnamefont
  {Heyes}}, \bibinfo {author} {\bibfnamefont {D.}~\bibnamefont {Dini}},
  \bibinfo {author} {\bibfnamefont {L.}~\bibnamefont {Costigliola}}, \ and\
  \bibinfo {author} {\bibfnamefont {J.~C.}\ \bibnamefont {Dyre}},\ }\bibfield
  {title} {\enquote {\bibinfo {title} {Transport coefficients of the
  {L}ennard-{J}ones fluid close to the freezing line},}\ }\href {\doibase
  10.1063/1.5128707} {\bibfield  {journal} {\bibinfo  {journal} {J. Chem.
  Phys.}\ }\textbf {\bibinfo {volume} {151}},\ \bibinfo {pages} {204502}
  (\bibinfo {year} {2019})}\BibitemShut {NoStop}%
\bibitem [{\citenamefont {Rosenfeld}(1976{\natexlab{a}})}]{RosenfeldCPL1976}%
  \BibitemOpen
  \bibfield  {author} {\bibinfo {author} {\bibfnamefont {Y.}~\bibnamefont
  {Rosenfeld}},\ }\bibfield  {title} {\enquote {\bibinfo {title} {Additivity of
  melting curves},}\ }\href {\doibase 10.1016/0009-2614(76)80048-6} {\bibfield
  {journal} {\bibinfo  {journal} {Chem. Phys. Lett.}\ }\textbf {\bibinfo
  {volume} {38}},\ \bibinfo {pages} {591--593} (\bibinfo {year}
  {1976}{\natexlab{a}})}\BibitemShut {NoStop}%
\bibitem [{\citenamefont
  {Rosenfeld}(1976{\natexlab{b}})}]{RosenfeldMolPhys1976}%
  \BibitemOpen
  \bibfield  {author} {\bibinfo {author} {\bibfnamefont {Y.}~\bibnamefont
  {Rosenfeld}},\ }\bibfield  {title} {\enquote {\bibinfo {title} {Universality
  of melting and freezing indicators and additivity of melting curves},}\
  }\href {\doibase 10.1080/00268977600102381} {\bibfield  {journal} {\bibinfo
  {journal} {Mol. Phys.}\ }\textbf {\bibinfo {volume} {32}},\ \bibinfo {pages}
  {963--977} (\bibinfo {year} {1976}{\natexlab{b}})}\BibitemShut {NoStop}%
\bibitem [{\citenamefont {Khrapak}\ \emph {et~al.}(2010)\citenamefont
  {Khrapak}, \citenamefont {Chaudhuri},\ and\ \citenamefont
  {Morfill}}]{KhrapakPRB2010}%
  \BibitemOpen
  \bibfield  {author} {\bibinfo {author} {\bibfnamefont {S.~A.}\ \bibnamefont
  {Khrapak}}, \bibinfo {author} {\bibfnamefont {M.}~\bibnamefont {Chaudhuri}},
  \ and\ \bibinfo {author} {\bibfnamefont {G.~E.}\ \bibnamefont {Morfill}},\
  }\bibfield  {title} {\enquote {\bibinfo {title} {Liquid-solid phase
  transition in the {L}ennard-{J}ones system},}\ }\href {\doibase
  10.1103/physrevb.82.052101} {\bibfield  {journal} {\bibinfo  {journal} {Phys.
  Rev. B}\ }\textbf {\bibinfo {volume} {82}},\ \bibinfo {pages} {052101}
  (\bibinfo {year} {2010})}\BibitemShut {NoStop}%
\bibitem [{\citenamefont {Khrapak}\ and\ \citenamefont
  {Morfill}(2011)}]{KhrapakJCP2011_2}%
  \BibitemOpen
  \bibfield  {author} {\bibinfo {author} {\bibfnamefont {S.~A.}\ \bibnamefont
  {Khrapak}}\ and\ \bibinfo {author} {\bibfnamefont {G.~E.}\ \bibnamefont
  {Morfill}},\ }\bibfield  {title} {\enquote {\bibinfo {title} {Accurate
  freezing and melting equations for the {L}ennard-{J}ones system},}\ }\href
  {\doibase 10.1063/1.3561698} {\bibfield  {journal} {\bibinfo  {journal} {J.
  Chem. Phys.}\ }\textbf {\bibinfo {volume} {134}},\ \bibinfo {pages} {094108}
  (\bibinfo {year} {2011})}\BibitemShut {NoStop}%
\bibitem [{\citenamefont {Pedersen}\ \emph {et~al.}(2016)\citenamefont
  {Pedersen}, \citenamefont {Costigliola}, \citenamefont {Bailey},
  \citenamefont {Schr{\o}der},\ and\ \citenamefont
  {Dyre}}]{PedersenNatCom2016}%
  \BibitemOpen
  \bibfield  {author} {\bibinfo {author} {\bibfnamefont {U.~R.}\ \bibnamefont
  {Pedersen}}, \bibinfo {author} {\bibfnamefont {L.}~\bibnamefont
  {Costigliola}}, \bibinfo {author} {\bibfnamefont {N.~P.}\ \bibnamefont
  {Bailey}}, \bibinfo {author} {\bibfnamefont {T.~B.}\ \bibnamefont
  {Schr{\o}der}}, \ and\ \bibinfo {author} {\bibfnamefont {J.~C.}\ \bibnamefont
  {Dyre}},\ }\bibfield  {title} {\enquote {\bibinfo {title} {Thermodynamics of
  freezing and melting},}\ }\href {\doibase 10.1038/ncomms12386} {\bibfield
  {journal} {\bibinfo  {journal} {Nature Commun.}\ }\textbf {\bibinfo {volume}
  {7}},\ \bibinfo {pages} {12386} (\bibinfo {year} {2016})}\BibitemShut
  {NoStop}%
\bibitem [{\citenamefont {Costigliola}\ \emph {et~al.}(2016)\citenamefont
  {Costigliola}, \citenamefont {Schr{\o}der},\ and\ \citenamefont
  {Dyre}}]{CostigliolaPCCP2016}%
  \BibitemOpen
  \bibfield  {author} {\bibinfo {author} {\bibfnamefont {L.}~\bibnamefont
  {Costigliola}}, \bibinfo {author} {\bibfnamefont {T.~B.}\ \bibnamefont
  {Schr{\o}der}}, \ and\ \bibinfo {author} {\bibfnamefont {J.~C.}\ \bibnamefont
  {Dyre}},\ }\bibfield  {title} {\enquote {\bibinfo {title} {Freezing and
  melting line invariants of the {L}ennard-{J}ones system},}\ }\href {\doibase
  10.1039/c5cp06363a} {\bibfield  {journal} {\bibinfo  {journal} {Phys. Chem.
  Chem. Phys.}\ }\textbf {\bibinfo {volume} {18}},\ \bibinfo {pages}
  {14678--14690} (\bibinfo {year} {2016})}\BibitemShut {NoStop}%
\bibitem [{\citenamefont {Khrapak}(2022{\natexlab{b}})}]{KhrapakJCP2022}%
  \BibitemOpen
  \bibfield  {author} {\bibinfo {author} {\bibfnamefont {S.~A.}\ \bibnamefont
  {Khrapak}},\ }\bibfield  {title} {\enquote {\bibinfo {title} {Gas-liquid
  crossover in the {L}ennard-{J}ones system},}\ }\href {\doibase
  10.1063/5.0085181} {\bibfield  {journal} {\bibinfo  {journal} {J. Chem.
  Phys.}\ }\textbf {\bibinfo {volume} {156}},\ \bibinfo {pages} {116101}
  (\bibinfo {year} {2022}{\natexlab{b}})}\BibitemShut {NoStop}%
\bibitem [{\citenamefont {Rosenfeld}(1999)}]{RosenfeldJPCM1999}%
  \BibitemOpen
  \bibfield  {author} {\bibinfo {author} {\bibfnamefont {Y.}~\bibnamefont
  {Rosenfeld}},\ }\bibfield  {title} {\enquote {\bibinfo {title} {A
  quasi-universal scaling law for atomic transport in simple fluids},}\ }\href
  {\doibase 10.1088/0953-8984/11/28/303} {\bibfield  {journal} {\bibinfo
  {journal} {J. Phys.: Condens. Matter}\ }\textbf {\bibinfo {volume} {11}},\
  \bibinfo {pages} {5415--5427} (\bibinfo {year} {1999})}\BibitemShut {NoStop}%
\bibitem [{\citenamefont {Rosenfeld}(1977)}]{RosenfeldPRA1977}%
  \BibitemOpen
  \bibfield  {author} {\bibinfo {author} {\bibfnamefont {Y.}~\bibnamefont
  {Rosenfeld}},\ }\bibfield  {title} {\enquote {\bibinfo {title} {Relation
  between the transport coefficients and the internal entropy of simple
  systems},}\ }\href {\doibase 10.1103/physreva.15.2545} {\bibfield  {journal}
  {\bibinfo  {journal} {Phys. Rev. A}\ }\textbf {\bibinfo {volume} {15}},\
  \bibinfo {pages} {2545--2549} (\bibinfo {year} {1977})}\BibitemShut {NoStop}%
\bibitem [{\citenamefont {Khrapak}(2023, submitted)}]{KhrapakPR2023}%
  \BibitemOpen
  \bibfield  {author} {\bibinfo {author} {\bibfnamefont {S.~A.}\ \bibnamefont
  {Khrapak}},\ }\bibfield  {title} {\enquote {\bibinfo {title} {Elementary
  vibrational model for transport properties of dense fluids},}\ }\href@noop {}
  {\bibfield  {journal} {\bibinfo  {journal} {Phys. Rep.}\ } (\bibinfo {year}
  {2023, submitted})}\BibitemShut {NoStop}%
\bibitem [{\citenamefont {Zwanzig}\ and\ \citenamefont
  {Mountain}(1965)}]{ZwanzigJCP1965}%
  \BibitemOpen
  \bibfield  {author} {\bibinfo {author} {\bibfnamefont {R.}~\bibnamefont
  {Zwanzig}}\ and\ \bibinfo {author} {\bibfnamefont {R.~D.}\ \bibnamefont
  {Mountain}},\ }\bibfield  {title} {\enquote {\bibinfo {title} {High-frequency
  elastic moduli of simple fluids},}\ }\href {\doibase 10.1063/1.1696718}
  {\bibfield  {journal} {\bibinfo  {journal} {J. Chem. Phys.}\ }\textbf
  {\bibinfo {volume} {43}},\ \bibinfo {pages} {4464--4471} (\bibinfo {year}
  {1965})}\BibitemShut {NoStop}%
\bibitem [{\citenamefont {Khrapak}(2020)}]{KhrapakMolecules2020}%
  \BibitemOpen
  \bibfield  {author} {\bibinfo {author} {\bibfnamefont {S.~A.}\ \bibnamefont
  {Khrapak}},\ }\bibfield  {title} {\enquote {\bibinfo {title} {Sound
  velocities of {L}ennard-{J}ones systems near the liquid-solid phase
  transition},}\ }\href {\doibase 10.3390/molecules25153498} {\bibfield
  {journal} {\bibinfo  {journal} {Molecules}\ }\textbf {\bibinfo {volume}
  {25}},\ \bibinfo {pages} {3498} (\bibinfo {year} {2020})}\BibitemShut
  {NoStop}%
\bibitem [{\citenamefont {Stephan}\ \emph {et~al.}(2019)\citenamefont
  {Stephan}, \citenamefont {Thol}, \citenamefont {Vrabec},\ and\ \citenamefont
  {Hasse}}]{Stephan2019}%
  \BibitemOpen
  \bibfield  {author} {\bibinfo {author} {\bibfnamefont {S.}~\bibnamefont
  {Stephan}}, \bibinfo {author} {\bibfnamefont {M.}~\bibnamefont {Thol}},
  \bibinfo {author} {\bibfnamefont {J.}~\bibnamefont {Vrabec}}, \ and\ \bibinfo
  {author} {\bibfnamefont {H.}~\bibnamefont {Hasse}},\ }\bibfield  {title}
  {\enquote {\bibinfo {title} {Thermophysical properties of the
  {L}ennard-{J}ones fluid: {D}atabase and data assessment},}\ }\href {\doibase
  10.1021/acs.jcim.9b00620} {\bibfield  {journal} {\bibinfo  {journal} {J.
  Chem. Informat. Model.}\ }\textbf {\bibinfo {volume} {59}},\ \bibinfo {pages}
  {4248--4265} (\bibinfo {year} {2019})}\BibitemShut {NoStop}%
\bibitem [{\citenamefont {Stephan}\ \emph {et~al.}(2020)\citenamefont
  {Stephan}, \citenamefont {Staubach},\ and\ \citenamefont
  {Hasse}}]{Stephan2020}%
  \BibitemOpen
  \bibfield  {author} {\bibinfo {author} {\bibfnamefont {S.}~\bibnamefont
  {Stephan}}, \bibinfo {author} {\bibfnamefont {J.}~\bibnamefont {Staubach}}, \
  and\ \bibinfo {author} {\bibfnamefont {H.}~\bibnamefont {Hasse}},\ }\bibfield
   {title} {\enquote {\bibinfo {title} {Review and comparison of equations of
  state for the {L}ennard-{J}ones fluid},}\ }\href {\doibase
  10.1016/j.fluid.2020.112772} {\bibfield  {journal} {\bibinfo  {journal}
  {Fluid Phase Equilibria}\ }\textbf {\bibinfo {volume} {523}},\ \bibinfo
  {pages} {112772} (\bibinfo {year} {2020})}\BibitemShut {NoStop}%
\bibitem [{\citenamefont {Thol}\ \emph {et~al.}(2016)\citenamefont {Thol},
  \citenamefont {Rutkai}, \citenamefont {K\"{o}ster}, \citenamefont {Lustig},
  \citenamefont {Span},\ and\ \citenamefont {Vrabec}}]{Thol2016}%
  \BibitemOpen
  \bibfield  {author} {\bibinfo {author} {\bibfnamefont {M.}~\bibnamefont
  {Thol}}, \bibinfo {author} {\bibfnamefont {G.}~\bibnamefont {Rutkai}},
  \bibinfo {author} {\bibfnamefont {A.}~\bibnamefont {K\"{o}ster}}, \bibinfo
  {author} {\bibfnamefont {R.}~\bibnamefont {Lustig}}, \bibinfo {author}
  {\bibfnamefont {R.}~\bibnamefont {Span}}, \ and\ \bibinfo {author}
  {\bibfnamefont {J.}~\bibnamefont {Vrabec}},\ }\bibfield  {title} {\enquote
  {\bibinfo {title} {Equation of state for the {L}ennard-{J}ones fluid},}\
  }\href {\doibase 10.1063/1.4945000} {\bibfield  {journal} {\bibinfo
  {journal} {J. Phys. Chem. Ref. Data}\ }\textbf {\bibinfo {volume} {45}},\
  \bibinfo {pages} {023101} (\bibinfo {year} {2016})}\BibitemShut {NoStop}%
\bibitem [{\citenamefont {Nasrabad}\ \emph {et~al.}(2006)\citenamefont
  {Nasrabad}, \citenamefont {Laghaei},\ and\ \citenamefont
  {Eu}}]{NasrabadJCP2006}%
  \BibitemOpen
  \bibfield  {author} {\bibinfo {author} {\bibfnamefont {A.~E.}\ \bibnamefont
  {Nasrabad}}, \bibinfo {author} {\bibfnamefont {R.}~\bibnamefont {Laghaei}}, \
  and\ \bibinfo {author} {\bibfnamefont {B.~C.}\ \bibnamefont {Eu}},\
  }\bibfield  {title} {\enquote {\bibinfo {title} {Molecular theory of thermal
  conductivity of the {L}ennard-{J}ones fluid},}\ }\href {\doibase
  10.1063/1.2166394} {\bibfield  {journal} {\bibinfo  {journal} {J. Chem.
  Phys.}\ }\textbf {\bibinfo {volume} {124}},\ \bibinfo {pages} {084506}
  (\bibinfo {year} {2006})}\BibitemShut {NoStop}%
\bibitem [{\citenamefont {Galliero}\ and\ \citenamefont
  {Boned}(2009)}]{GallieroPRE2009}%
  \BibitemOpen
  \bibfield  {author} {\bibinfo {author} {\bibfnamefont {G.}~\bibnamefont
  {Galliero}}\ and\ \bibinfo {author} {\bibfnamefont {C.}~\bibnamefont
  {Boned}},\ }\bibfield  {title} {\enquote {\bibinfo {title} {Thermal
  conductivity of the {L}ennard-{J}ones chain fluid model},}\ }\href {\doibase
  10.1103/physreve.80.061202} {\bibfield  {journal} {\bibinfo  {journal} {Phys.
  Rev. E}\ }\textbf {\bibinfo {volume} {80}},\ \bibinfo {pages} {061202}
  (\bibinfo {year} {2009})}\BibitemShut {NoStop}%
\bibitem [{\citenamefont {Heyes}\ \emph {et~al.}(2023)\citenamefont {Heyes},
  \citenamefont {Dini}, \citenamefont {Pieprzyk},\ and\ \citenamefont
  {Bra{\'{n}}ka}}]{HeyesJCP2023}%
  \BibitemOpen
  \bibfield  {author} {\bibinfo {author} {\bibfnamefont {D.~M.}\ \bibnamefont
  {Heyes}}, \bibinfo {author} {\bibfnamefont {D.}~\bibnamefont {Dini}},
  \bibinfo {author} {\bibfnamefont {S.}~\bibnamefont {Pieprzyk}}, \ and\
  \bibinfo {author} {\bibfnamefont {A.~C.}\ \bibnamefont {Bra{\'{n}}ka}},\
  }\bibfield  {title} {\enquote {\bibinfo {title} {Departures from perfect
  isomorph behavior in {L}ennard-{J}ones fluids and solids},}\ }\href {\doibase
  10.1063/5.0143651} {\bibfield  {journal} {\bibinfo  {journal} {J. Chem.
  Phys.}\ }\textbf {\bibinfo {volume} {158}},\ \bibinfo {pages} {134502}
  (\bibinfo {year} {2023})}\BibitemShut {NoStop}%
\bibitem [{\citenamefont {Hirschfelder}\ \emph {et~al.}(1954)\citenamefont
  {Hirschfelder}, \citenamefont {Curtiss},\ and\ \citenamefont
  {Bird}}]{HirschfelderBook}%
  \BibitemOpen
  \bibfield  {author} {\bibinfo {author} {\bibfnamefont {J.~O.}\ \bibnamefont
  {Hirschfelder}}, \bibinfo {author} {\bibfnamefont {C.~F.}\ \bibnamefont
  {Curtiss}}, \ and\ \bibinfo {author} {\bibfnamefont {R.~B.}\ \bibnamefont
  {Bird}},\ }\href@noop {} {\emph {\bibinfo {title} {The Molecular Theory of
  Gases and Liquids -}}}\ (\bibinfo  {publisher} {Wiley},\ \bibinfo {address}
  {New York},\ \bibinfo {year} {1954})\BibitemShut {NoStop}%
\bibitem [{\citenamefont {Hirschfelder}\ \emph {et~al.}(1948)\citenamefont
  {Hirschfelder}, \citenamefont {Bird},\ and\ \citenamefont
  {Spotz}}]{HirschfelderJCP1948}%
  \BibitemOpen
  \bibfield  {author} {\bibinfo {author} {\bibfnamefont {J.~O.}\ \bibnamefont
  {Hirschfelder}}, \bibinfo {author} {\bibfnamefont {R.~B.}\ \bibnamefont
  {Bird}}, \ and\ \bibinfo {author} {\bibfnamefont {E.~L.}\ \bibnamefont
  {Spotz}},\ }\bibfield  {title} {\enquote {\bibinfo {title} {The transport
  properties for non-polar gases},}\ }\href {\doibase 10.1063/1.1746696}
  {\bibfield  {journal} {\bibinfo  {journal} {J. Chem. Phys.}\ }\textbf
  {\bibinfo {volume} {16}},\ \bibinfo {pages} {968--981} (\bibinfo {year}
  {1948})}\BibitemShut {NoStop}%
\bibitem [{\citenamefont {Smith}\ and\ \citenamefont
  {Munn}(1964)}]{SmithJCP1964}%
  \BibitemOpen
  \bibfield  {author} {\bibinfo {author} {\bibfnamefont {F.~J.}\ \bibnamefont
  {Smith}}\ and\ \bibinfo {author} {\bibfnamefont {R.~J.}\ \bibnamefont
  {Munn}},\ }\bibfield  {title} {\enquote {\bibinfo {title} {Automatic
  calculation of the transport collision integrals with tables for the {M}orse
  potential},}\ }\href {\doibase 10.1063/1.1725768} {\bibfield  {journal}
  {\bibinfo  {journal} {J. Chem. Phys.}\ }\textbf {\bibinfo {volume} {41}},\
  \bibinfo {pages} {3560--3568} (\bibinfo {year} {1964})}\BibitemShut {NoStop}%
\bibitem [{\citenamefont
  {Khrapak}(2014{\natexlab{a}})}]{KhrapakPRE2014_scattering}%
  \BibitemOpen
  \bibfield  {author} {\bibinfo {author} {\bibfnamefont {S.~A.}\ \bibnamefont
  {Khrapak}},\ }\bibfield  {title} {\enquote {\bibinfo {title} {Classical
  scattering in strongly attractive potentials},}\ }\href {\doibase
  10.1103/physreve.89.032145} {\bibfield  {journal} {\bibinfo  {journal} {Phys.
  Rev. E}\ }\textbf {\bibinfo {volume} {89}},\ \bibinfo {pages} {032145}
  (\bibinfo {year} {2014}{\natexlab{a}})}\BibitemShut {NoStop}%
\bibitem [{\citenamefont {Khrapak}(2014{\natexlab{b}})}]{KhrapakEPJD2014}%
  \BibitemOpen
  \bibfield  {author} {\bibinfo {author} {\bibfnamefont {S.~A.}\ \bibnamefont
  {Khrapak}},\ }\bibfield  {title} {\enquote {\bibinfo {title} {Accurate
  transport cross sections for the {L}ennard-{J}ones potential},}\ }\href
  {\doibase 10.1140/epjd/e2014-50449-y} {\bibfield  {journal} {\bibinfo
  {journal} {Eur. Phys. J. D}\ }\textbf {\bibinfo {volume} {68}},\ \bibinfo
  {pages} {276} (\bibinfo {year} {2014}{\natexlab{b}})}\BibitemShut {NoStop}%
\bibitem [{\citenamefont {Kim}\ and\ \citenamefont
  {Monroe}(2014)}]{KimJCompPhys2014}%
  \BibitemOpen
  \bibfield  {author} {\bibinfo {author} {\bibfnamefont {S.~U.}\ \bibnamefont
  {Kim}}\ and\ \bibinfo {author} {\bibfnamefont {C.~W.}\ \bibnamefont
  {Monroe}},\ }\bibfield  {title} {\enquote {\bibinfo {title} {High-accuracy
  calculations of sixteen collision integrals for {L}ennard-{J}ones (12-6)
  gases and their interpolation to parameterize neon, argon, and krypton},}\
  }\href {\doibase 10.1016/j.jcp.2014.05.018} {\bibfield  {journal} {\bibinfo
  {journal} {J. Comput. Phys.}\ }\textbf {\bibinfo {volume} {273}},\ \bibinfo
  {pages} {358--373} (\bibinfo {year} {2014})}\BibitemShut {NoStop}%
\bibitem [{\citenamefont {Kristiansen}(2020)}]{Kristiansen2020}%
  \BibitemOpen
  \bibfield  {author} {\bibinfo {author} {\bibfnamefont {K.~R.}\ \bibnamefont
  {Kristiansen}},\ }\bibfield  {title} {\enquote {\bibinfo {title} {Transport
  properties of the simple {L}ennard-{J}ones/spline fluid {I}: Binary
  scattering and high-accuracy low-density transport coefficients},}\ }\href
  {\doibase 10.3389/fphy.2020.00271} {\bibfield  {journal} {\bibinfo  {journal}
  {Frontiers Phys.}\ }\textbf {\bibinfo {volume} {8}},\ \bibinfo {pages} {271}
  (\bibinfo {year} {2020})}\BibitemShut {NoStop}%
\end{thebibliography}%

\end{document}